\documentclass[lettersize,journal]{IEEEtran}
\usepackage{cite}
\ifCLASSINFOpdf
  \usepackage[pdftex]{graphicx}
\else
  \usepackage[dvips]{graphicx}
\fi
\usepackage[cmex10]{amsmath}
\usepackage{amsmath,amsfonts,amssymb}
\usepackage{mdwmath}
\usepackage[ruled]{algorithm2e}
\usepackage{array}
\usepackage{caption}
\usepackage{textcomp}
\usepackage{fixltx2e}
\usepackage{bm}
\usepackage{stfloats}
\usepackage{url}
\usepackage{verbatim}
\usepackage{graphicx}
\usepackage{subfigure} 
\usepackage{mdwtab}
\usepackage{booktabs}
\usepackage{multirow}
\usepackage{etoolbox}
\usepackage{makecell}
\usepackage{xcolor}
\usepackage[bottom]{footmisc}
\usepackage{mdframed}
\usepackage{balance}
\usepackage{threeparttable}

\def\BibTeX{{\rm B\kern-.05em{\sc i\kern-.025em b}\kern-.08em
    T\kern-.1667em\lower.7ex\hbox{E}\kern-.125emX}}

\newcommand{\e}{\begin{equation}}
\newcommand{\ee}{\end{equation}}
\newcommand{\eqn}{\begin{eqnarray}}
\newcommand{\eeqn}{\end{eqnarray}}

\begin{document}
\vspace{-4mm}
 \title{Sensing User's Activity, Channel, and Location \\
 with {\color{black}Near-Field} Extra-Large-Scale MIMO}
\vspace{-2mm}
\author{Li Qiao, Anwen Liao, Zhuoran Li, Hua Wang, Zhen Gao, 
Xiang Gao, Yu Su, Pei Xiao,~\IEEEmembership{Senior Member,~IEEE},  \\
Li You,~\IEEEmembership{Senior Member,~IEEE}, and Derrick Wing Kwan Ng,~\IEEEmembership{Fellow,~IEEE}

 \vspace{-2mm}

\thanks{The work was supported in part by the National Key Research and Development Program of China under Grant 2020YFB1807900. The work of L. Qiao was supported by the China Scholarship Council. The work of Z. Gao was supported in part by the National Natural Science Foundation of China (NSFC) under Grant U2233216 and Grant 62071044, in part by the Shandong Province Natural Science Foundation under Grant ZR2022YQ62, and in part by the Beijing Nova Program. X. Gao is supported by NSFC under Grant 62203048. D. W. K. Ng is supported by the Australian Research Council's Discovery Projects (DP210102169, DP230100603). An earlier version of this paper was presented in part at the 2023 IEEE Wireless Communications and Networking Conference Workshops (WCNCW) \cite{WCNC2023}. [DOI:
10.1109/WCNC55385.2023.10118760].}
\thanks{L. Qiao, Z. Li, H. Wang, Z. Gao, and X. Gao are with Beijing Institute of Technology, Beijing 100081, China (e-mails: \{qiaoli, gaozhen16\}@bit.edu.cn).}
\thanks{A. Liao is with the Nanjing Research Institute of Electronics Technology, Nanjing 210039, China (e-mail: liaoanwen18@qq.com)}
\thanks{Y. Su is with China Mobile (Chengdu) Institute of Research and Development, Chengdu, Sichuan, 610000, China.  (e-mail: suyu@cmii.chinamobile.com)}
\thanks{L. Qiao and P. Xiao are with 5GIC \& 6GIC, the Institute for Communication Systems (ICS), University of Surrey, GU2 7XH Guildford, U.K. (e-mail: \{l.qiao, p.xiao\}@surrey.ac.uk).}
\thanks{L. You is with the National Mobile Communications Research Laboratory, Southeast University, Nanjing 210096, China, and also with the Purple Mountain Laboratories, Nanjing 211100, China (e-mail: \mbox{lyou@seu.edu.cn}).}
\thanks{D. W. K. Ng is with the School of Electrical Engineering and Telecommunications, University of New South Wales, Sydney,
NSW 2025, Australia (email: w.k.ng@unsw.edu.au).}
}

\maketitle
\begin{abstract}
This paper proposes a grant-free massive access scheme based on the millimeter wave (mmWave) extra-large-scale multiple-input multiple-output (XL-MIMO) to support massive Internet-of-Things (IoT) devices with low latency, high data rate, and high localization accuracy in the upcoming {\color{black}sixth-generation (6G) networks. The XL-MIMO consists of multiple antenna subarrays that are widely spaced over the service area to ensure line-of-sight (LoS) transmissions. First, we establish the XL-MIMO-based massive access model considering the near-field spatial non-stationary (SNS) property. Then, by exploiting the block sparsity of subarrays and the SNS property, we propose a structured block orthogonal matching pursuit algorithm for efficient active user detection (AUD) and channel estimation (CE).} Furthermore, different sensing matrices are applied in different pilot subcarriers for exploiting the diversity gains. Additionally, a multi-subarray collaborative localization algorithm is designed for localization. In particular, the angle of arrival (AoA) and time difference of arrival (TDoA) of the LoS links between active users and related subarrays are extracted from the estimated XL-MIMO channels, and then the coordinates of active users are acquired by jointly utilizing the AoAs and TDoAs. Simulation results show that the proposed algorithms outperform existing algorithms in terms of AUD and CE performance and can achieve centimeter-level localization accuracy.

\end{abstract}

\IEEEpeerreviewmaketitle

\section{Introduction}\label{S1}

The upcoming sixth-generation (6G) communication networks are anticipated to outperform the current fifth-generation (5G) networks substantially in all aspects, such as data rate, connection density, latency, and localization accuracy \cite{6G}. {\color{black}These improvements are necessary to satisfy the ever-stringent demands of emerging applications, such as Industry 4.0 \cite{Aceto_CST19}, cloud virtual reality \cite{CC-VR20}, and collaborative intelligence \cite{LI_FL}.} However, with the explosive growth of Internet-of-Things (IoT) devices, conventional grant-based random access protocols may lead to excessively large amount of signaling overhead and long latency when applied to IoT devices \cite{ChenXM_JSAC21}. Therefore, the industry and academic communities have recently explored grant-free random access protocols, e.g., \cite{Qiao2, 3GPP}. In addition to the demand of low-latency massive access, there is also a skyrocketed requirement for high data rates. To this end, millimeter-wave (mmWave) and even terahertz (THz) technologies with an abundance of unlicensed spectral resources are considered as promising solutions \cite{6G}. Meanwhile, benefitting from the small antenna size of high signal carrier frequency, deploying an extremely large number of antennas, which is also known as extra-large-scale multiple-input multiple-output (XL-MIMO), in a limited area to greatly improve the system spectral efficiency is possible \cite{Heath_WC20}. Moreover, due to their inherent characteristics of large bandwidth and array aperture, mmWave XL-MIMO systems naturally possess an enormous high-precision localization potential that remains unexploited \cite{2018_CST_localization}.

\subsection{Related Works}\label{S1.1}
 
In this subsection, we will introduce three areas of literature, i.e., grant-free random access protocols, the XL-MIMO technique, and cellular network-based localization.

As for grant-free random access, user devices (UDs) can directly transmit their signals without the need for dedicated scheduling by the base stations (BSs), leading to significant reduction in both the signaling overhead and latency \cite{ChenXM_JSAC21}. However, due to the limited radio resources and the large number of UDs, active user detection (AUD), channel estimation (CE), and data detection at the BSs are challenging to perform. Yet, thanks to the sporadic traffic of IoT devices, compressive sensing (CS)-based algorithms, including Bayesian-type \cite{Shim_Tcom19,YuanXJ_TWC20,YuW_TSP18_P1,KeML_TSP20,YuanXJ_TWC22, ChenW_TWC22, mmW-GF-gao2, IimoriH_TWC22, KeML_JSAC21,MYK}, greedy-type \cite{ DuY_WCL, WangY_IoTJ22, WuL_IoTJ21, Qiao_TVT}, optimization-type \cite{ShaoXD_TSP20,DjelouatH_WCL21, mmW-GF-shao1}, deep learning (DL)-based \cite{ChenXM_CL20, MaoZ_IoTJ22,Shim_IoTJ22} algorithms etc., have been widely studied for enabling efficient AUD, CE, and data detection in grant-free massive access. 

Specifically, considering single-antenna BSs, the authors of \cite{Shim_Tcom19} proposed an expectation propagation (EP)-based joint AUD and CE algorithm with relatively low computational complexity. Furthermore, an improved approximate message passing (AMP) algorithm was proposed in \cite{YuanXJ_TWC20} to realize the joint AUD, CE, and data detection. Besides, to further improve the AUD and CE performance, massive MIMO was adopted at the BSs in \cite{YuW_TSP18_P1,KeML_TSP20,YuanXJ_TWC22, ChenW_TWC22, KeML_JSAC21,mmW-GF-gao2} for massive connectivity under various practical settings. Specifically, the authors of \cite{YuW_TSP18_P1} designed a minimum mean square error (MMSE)-based vector AMP algorithm adopting the statistical channel information. Different from the narrow-band-based massive access schemes as in \cite{Shim_Tcom19, YuanXJ_TWC20, YuW_TSP18_P1}, broadband massive access and orthogonal frequency division multiplexing (OFDM) were considered in \cite{KeML_TSP20,YuanXJ_TWC22, ChenW_TWC22,KeML_JSAC21, mmW-GF-gao2} to provide throughput improvement. In particular, by exploiting the structured sparsity of massive MIMO channels in both the spatial and angular domains, a distributed CS (DCS) theory-based generalized multiple measurement vector AMP (GMMV-AMP) algorithm was proposed in \cite{KeML_TSP20} to jointly perform AUD and CE. Furthermore, a sparse Bayesian learning (SBL)-based algorithm was developed in \cite{ChenW_TWC22}, where the simultaneously row-sparse and clustered sparse structure of massive MIMO channels in the angular domain was utilized for the enhanced AUD and CE performance. Besides, by exploiting the correlations among OFDM subcarriers, the authors of \cite{YuanXJ_TWC22} proposed a turbo-type message passing algorithm for improving the joint AUD and CE performance. Furthermore, the authors of \cite{KeML_JSAC21} proposed the cloud computing and edge computing paradigms for cell-free massive access systems, where multiple access points cooperate with each other for performance improvement. 
Moreover, to further increase the data rate for enabling enhanced mobile broadband applications, mmWave/THz bands were adopted in \cite{mmW-GF-gao2} and XL-MIMO arrays were adopted in \cite{IimoriH_TWC22}, \cite{MYK}. Specifically, an orthogonal approximate message passing (OAMP)-based algorithm was proposed for joint AUD and CE in \cite{mmW-GF-gao2}. By modeling the channel with a nested Bernoulli-Gaussian distribution under the plane-wave hypothesis, the authors of \cite{IimoriH_TWC22} proposed a bilinear Bayesian-based joint AUD and CE algorithm. In addition, a geometric channel model exploiting spherical wavefronts was adopted for XL-MIMO in \cite{MYK} considering the use of mixed analog-to-digital converters, and an OAMP-based algorithm was proposed for joint AUD and CE. 
 
\begin{table*}[!tp]
\renewcommand\arraystretch{1.6}
\captionsetup{font = {footnotesize}, labelsep = newline} 
\caption{Comparison of different grant-free massive IoT access schemes}
\vspace{-1.0mm}
\centering
\label{TAB1}
\begin{threeparttable}\tiny
\begin{tabular}{|p{2cm}<{\centering}|p{1.2cm}<{\centering}|p{2.8cm}<{\centering}|c|p{2.8cm}<{\centering}|p{2.8cm}<{\centering}|} 
\hline
 Categories & References & Array type at BS & Channel model & Frequency band \& bandwidth & Algorithm \\
\hline
 \multirow{6}{*}{\makecell{Bayesian algorithms}} & \cite{Shim_Tcom19}/ \cite{YuanXJ_TWC20} & Single antenna & Rayleigh channel & Low-frequency \& narrow-band & EP/ AMP \\
\cline{2-6}
  & \cite{YuW_TSP18_P1} & Massive MIMO & Rayleigh channel & Low-frequency \& narrow-band & Vector AMP \\
\cline{2-6}
  & \cite{KeML_TSP20}/ \cite{ChenW_TWC22} & Massive MIMO & Geometric multipath channel & Low-frequency \& wideband & AMP/ SBL \\
\cline{2-6}
  & \cite{YuanXJ_TWC22} & MIMO & Rayleigh channel & Low-frequency \& wideband & Turbo-MP \\
\cline{2-6}
  & \cite{KeML_JSAC21} & Cell-free massive MIMO & Geometric multipath channel & Low-frequency \& wideband & AMP \\
\cline{2-6}
& \cite{mmW-GF-gao2} & Massive MIMO & Geometric multipath channel & \makecell{mmWave or THz \& wideband} & OAMP \\
\cline{2-6}
& \cite{IimoriH_TWC22}/ \cite{MYK} & XL-MIMO & \makecell{Nested Bernoulli-Gaussian channel/\\  Geometric spherical wavefronts} & Low-frequency \& narrow-band & Bilinear Bayesian/ OAMP \\

\hline
 \multirow{3}{*}{\makecell{Optimization algorithms}} & \cite{ShaoXD_TSP20} & Massive MIMO & Rayleigh channel & Low-frequency \& narrow-band & \makecell{Riemannian optimization} \\
\cline{2-6}
  & \cite{DjelouatH_WCL21} & Massive MIMO & Geometric multipath channel & Low-frequency \& narrow-band & ADMM \\
\cline{2-6}
& \cite{mmW-GF-shao1} & Massive MIMO & Geometric multipath channel & \makecell{mmWave or THz \& wideband} & \makecell{Low-rank optimization} \\
\hline
 \multirow{3}{*}{\makecell{DL-based algorithms}} & \cite{ChenXM_CL20} & Massive MIMO & Rayleigh channel & Low-frequency \& narrow-band & \makecell{AMP-based DL} \\
\cline{2-6}
  & \cite{MaoZ_IoTJ22} & Single antenna or MIMO & Rayleigh channel & Low-frequency \& narrow-band & \makecell{ADMM-based DL} \\
\cline{2-6}
  & \cite{Shim_IoTJ22} & Single antenna & Rayleigh channel & Low-frequency \& wideband & LSTM framework \\
\hline
 \multirow{5}{*}{\makecell{Greedy algorithms}} & \cite{DuY_WCL}  & Single antenna & Rayleigh channel & Low-frequency \& narrow-band & BSP \\
\cline{2-6}
  & \cite{WangY_IoTJ22} & Single antenna & Rayleigh channel & Low-frequency \& narrow-band & BOMP-based  \\
\cline{2-6}
  & \cite{WuL_IoTJ21} & Massive MIMO & Rayleigh channel & Low-frequency \& narrow-band & STS-ASP \\
\cline{2-6}
  & \cite{Qiao_TVT} & Massive MIMO & Rayleigh channel & Low-frequency \& narrow-band & StrOMP \& SIC-SSP \\
\cline{2-6}
  & \makecell{Our work} & \makecell{XL-MIMO\\(Multiple subarrays)} & \makecell{Geometric multipath channel \& \\ Near-field SNS property} & \makecell{mmWave \& wideband} & \makecell{OMP-based} \\
\hline
\end{tabular}
\end{threeparttable}
\label{Compar}
\end{table*}

On the other hand, greedy-type \cite{DuY_WCL, WangY_IoTJ22, WuL_IoTJ21, Qiao_TVT} and optimization-type \cite{ShaoXD_TSP20,DjelouatH_WCL21, mmW-GF-shao1} CS algorithms are also effective for solving high-dimensional massive IoT access problems. Specifically, a block subspace pursuit (BSP)-based algorithm was proposed in \cite{DuY_WCL} for efficient joint AUD, CE, and data detection. Furthermore, by exploiting the noncontinuous temporal correlation of UDs, the authors of \cite{WangY_IoTJ22} proposed a block orthogonal matching pursuit (BOMP)-based algorithm for improving the performance. In addition, massive MIMO at the BSs was considered in \cite{WuL_IoTJ21, Qiao_TVT} for improving the performance of AUD. In particular, the authors of \cite{WuL_IoTJ21} proposed a spatial–temporal structure enhanced adaptive subspace pursuit (STS-ASP) for joint AUD and CE. Furthermore, a structured orthogonal matching pursuit (StrOMP) algorithm for activity detection and successive interference cancellation (SIC)-based structured subspace pursuit (SSP) algorithm for data detection were proposed in \cite{Qiao_TVT}.

As for optimization-based CS approaches for grant-free massive access \cite{ShaoXD_TSP20,DjelouatH_WCL21, mmW-GF-shao1}, by transforming the joint AUD and CE into a low-dimensional optimization problem with a full column rank constraint, a Riemannian trust-region based algorithm was proposed in \cite{ShaoXD_TSP20} to reduce the size of the search space of a multiple measurement vector (MMV)-CS problem. Furthermore, an alternating direction method of multipliers (ADMM)-based method was proposed in \cite{DjelouatH_WCL21} to solve the sparse MMV reconstruction problem by exploiting the second-order statistics of the spatially correlated channels. To further improve the data rate, mmWave/THz wideband massive access was considered in \cite{mmW-GF-shao1}, where two multi-rank aware joint AUD and CE algorithms were proposed by exploiting the low-rank and sparse characteristics in the delay-angular domain.

More recently, DL-based approaches have been applied to massive IoT access to achieve smart wireless communications \cite{ChenXM_CL20, MaoZ_IoTJ22,Shim_IoTJ22}. For instance, the authors in \cite{ChenXM_CL20} and \cite{MaoZ_IoTJ22} proposed the model-driven DL frameworks based on the AMP algorithm and the ADMM algorithm for massive grant-free random access, respectively. In addition, a long short-term memory (LSTM) framework-based AUD and CE scheme was proposed in \cite{Shim_IoTJ22}, where the DL network processes the direct mapping between the received signals and the indices of active UDs and the associated channels. For clarity, the comparison of different grant-free massive IoT access schemes \cite{KeML_JSAC21,MYK,IimoriH_TWC22, Shim_Tcom19,YuanXJ_TWC20,YuW_TSP18_P1,KeML_TSP20,YuanXJ_TWC22, ChenW_TWC22, mmW-GF-gao2, DuY_WCL, WangY_IoTJ22, WuL_IoTJ21, Qiao_TVT,ShaoXD_TSP20,DjelouatH_WCL21, mmW-GF-shao1,ChenXM_CL20, MaoZ_IoTJ22,Shim_IoTJ22} is summarized in Table \ref{Compar}.
 
It can be seen that most of the related works focus on low-frequency band, e.g., \cite{KeML_JSAC21,MYK,IimoriH_TWC22,Shim_Tcom19,YuanXJ_TWC20,YuW_TSP18_P1,
KeML_TSP20,YuanXJ_TWC22, ChenW_TWC22, DuY_WCL, WangY_IoTJ22, WuL_IoTJ21, Qiao_TVT,ShaoXD_TSP20,DjelouatH_WCL21, ChenXM_CL20, MaoZ_IoTJ22,Shim_IoTJ22}. {\color{black}To further improve the data rate for emerging IoT applications, e.g., federated edge learning \cite{LI_FL}, the authors of \cite{mmW-GF-gao2} and \cite{mmW-GF-shao1} considered mmWave/THz wideband massive access.} As the frequency increases, the number of antennas also increases from a single-antenna setting (e.g., \cite{Shim_IoTJ22,WangY_IoTJ22}) to massive MIMO (e.g., \cite{KeML_TSP20,mmW-GF-shao1}) and even XL-MIMO (i.e., \cite{IimoriH_TWC22, MYK}). XL-MIMO differs from massive MIMO mainly due to the presence of spherical wavefront and spatial non-stationary (SNS) property. Specifically, the Rayleigh distance, which is proportional to the square of the array aperture and inversely proportional to the carrier wavelength \cite{Stutzman_Antenna12}, indicates that mmWave communication systems employing XL-MIMO arrays may encounter near-field SNS channels \cite{Heath_WC20, MarinelloJC_TVT20}. As a result, different parts of the XL-MIMO array may also view different UDs and the near-field spherical wavefront needs to be considered. In fact, the traditional far-field signal plane wave hypothesis adopted in massive MIMO, e.g., \cite{KeML_TSP20,mmW-GF-shao1}, is no longer applicable, which motivates the investigation of near-field SNS XL-MIMO channel estimation, e.g., \cite{JinS_WCL20,ChenJ_WCL22,JinS_JSAC20,HouS_CL20,DaiLL_CL21,DaiLL_Tcom22}. To be specific, by utilizing a hierarchical-block prior to characterizing the delay-domain sparse structure of XL-MIMO channels, an iterative SBL scheme was proposed in \cite{ChenJ_WCL22} to infer the channel vectors and unknown hyperparameters.
 Also, the authors in \cite{JinS_JSAC20} regarded the near-field SNS channel as an image and proposed a model-driven deep learning-based downlink channel reconstruction scheme, which introduced the You-Only-Look-Once (YOLO) neural network to estimate the angles and delays.
 Furthermore, by exploiting the delay-domain sparsity of the SNS channels, a block matching pursuit-based two-stage sparse channel estimation scheme was proposed in \cite{HouS_CL20} to estimate the massive MIMO channels. Besides, the authors in \cite{DaiLL_CL21,DaiLL_Tcom22} exploited the polar-domain channel sparsity of near-field XL-MIMO channels and proposed several improved OMP algorithms to estimate the XL-MIMO channels. Moreover, the authors in \cite{DaiLL_CL21,DaiLL_Tcom22} analyzed the conversion between the near-field and far-field conditions and revealed the hybrid-field channel characteristics of the XL-MIMO channels. {\color{black}In addition, for intelligent reflecting surface (IRS)-assisted mmWave communication systems, a hybrid multi-objective evolutionary paradigm for channel estimation and a deep-unfolding-based algorithm for beamforming were proposed in \cite{Tangjie} and \cite{Yunlong}, respectively.} Motivated by \cite{JinS_WCL20,ChenJ_WCL22,JinS_JSAC20,HouS_CL20,DaiLL_CL21,DaiLL_Tcom22}, for XL-MIMO-based high-rate massive IoT access, exploiting the inherent properties of XL-MIMO channel is a promising approach to reducing the access latency, which has not been thoroughly investigated. 

Moreover, due to the large bandwidth and large array aperture in mmWave XL-MIMO systems, the resolutions of delay and angle are  high, which is promising for high-precision localization while performing massive IoT access.
\textcolor{black}{
Compared to satellite localization that is vulnerable to blockages, e.g., building occlusion, cellular network localization is more suitable for indoor environment \cite{2018_CST_localization}.
In the literature, traditional cellular network localization methods can be divided into three categories, depending on the receive signal strength (RSS), time of arrival (ToA)/time difference of arrival (TDoA), and angle of arrival (AoA)/angle of departure (AoD), respectively \cite{2018_CST_localization}. Specifically, the RSS-based localization is commonly used for indoor low-frequency rich scattering environments \cite{2018_TCOM_RSSindoor}.
However, the severe path loss in the mmWave band hinders its practical implementation. Furthermore, ToA-based methods estimate the delays between multiple anchors and the UD to obtain the UD’s location according to the intersection of several circles \cite{2017_TSP_DirectLoc_ToA, 2ToA1TDoA}, while TDoA-based methods resort to the time differences between the anchors and the UD to conduct hyperbolic localization \cite{2015_CC_TDoA}.
Compared with the ToA-based methods that require accurate time synchronization between the UD and the anchors \cite{2015_CC_TDoA}, the TDoA-based methods only require the anchors to be synchronized with each other, making them easier to implement. In addition, benefiting from high angle resolution of mMIMO systems, the AoA-based localization has been widely studied \cite{2017_TSP_DirectLoc_ToA,2018_TWC_AoA}. More recently, IRS was utilized in \cite{23JSTSP_Li} to help sense the UD's channel and location, where customized OMP algorithm, customized multiple
signal classification (MUSIC) algorithm, and polar-domain gradient descent algorithm were adopted. On the industrial front, 5G new radio (NR) Release 17 adopts several methods, e.g., TDoAs, round trip times (RTTs), and AoAs/AoDs, for localization through cooperation among several BSs \cite{ref_PositioningIn5GNetworks,ref_3GPP_38215}. However, most of the previous works studied the single-user localization scenarios, which leaves a lot of research space for joint multiple access and localization.
}
 
\begin{figure}[!tp]
\vspace{-5mm}
\begin{center}
 \includegraphics[width=0.75\columnwidth,keepaspectratio]{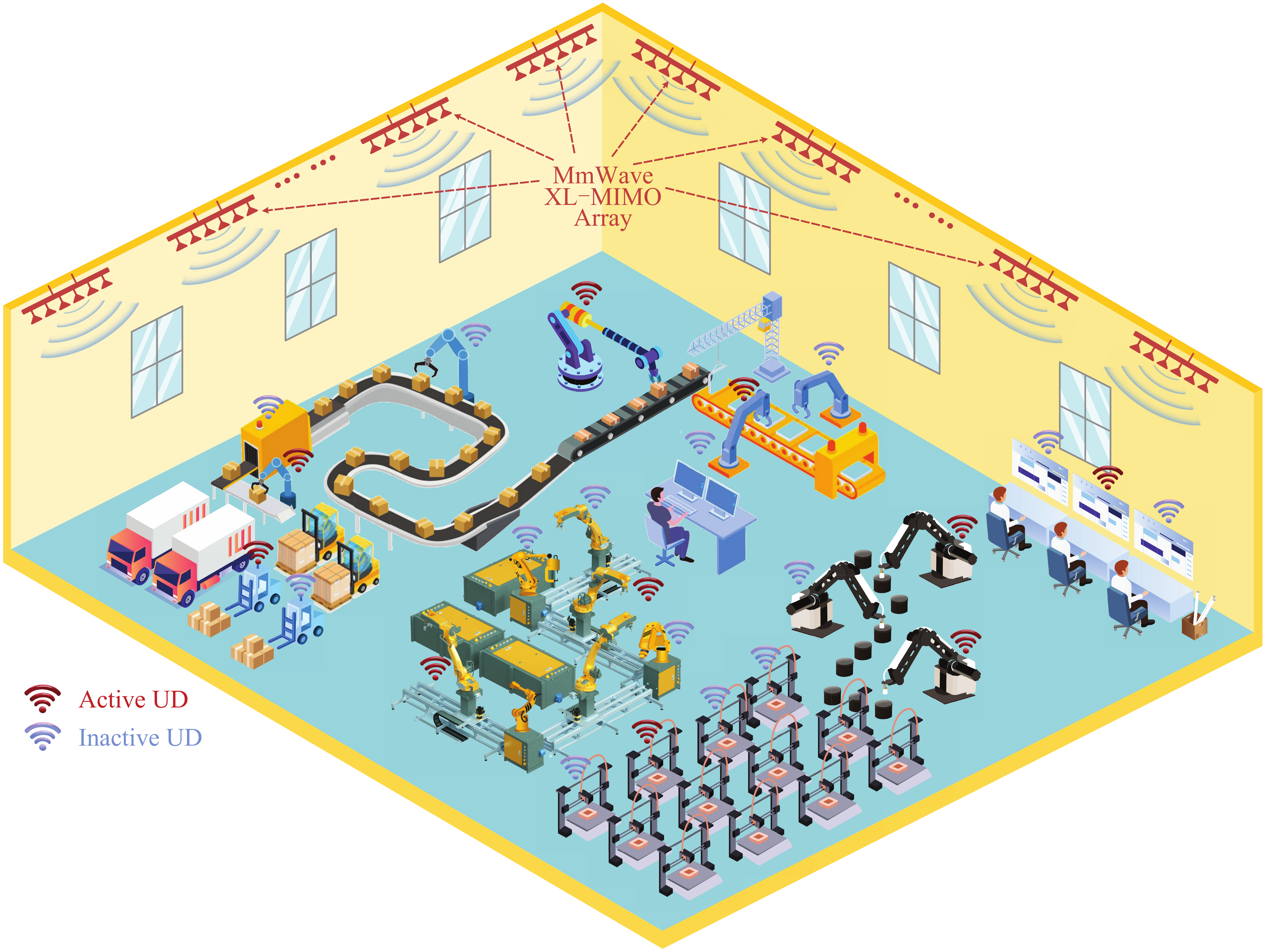}
\end{center}
\setlength{\abovecaptionskip}{-0.2mm}
 \captionsetup{font = {footnotesize}, name = {Fig.}, labelsep = period} 
\caption{Typical massive IoT access in smart factory scenarios, where the mmWave BS equipped with an XL-MIMO array can provide low-latency high data-rate uplink access and high-accuracy localization services for numerous UDs.}
 \label{FIG1}
  \vspace{-8mm}
\end{figure}

\subsection{Our Contributions}\label{S1.3}

{\color{black}Motivated by the above analysis, this paper investigates the grant-free massive IoT access and localization in indoor scenarios, where the mmWave XL-MIMO system with large bandwidth is adopted for improving the data rate as well as localization accuracy. To address the blockage and high attenuation of mmWave signals \cite{mmW-GF-shao1}, the mmWave XL-MIMO array equipped at the BS is split into multiple subarrays, as shown in Fig.~\ref{FIG1}. We first formulate the massive access signal model and the XL-MIMO channel model. Then, we propose a structured block OMP (StrBOMP) algorithm for efficient joint AUD and CE, which exploit the block sparsity of subarrays and the near-field SNS property of the XL-MIMO channel. Furthermore, we design GMMV sensing matrices for improving AUD and CE performance, where the GMMV setting outperforms MMV due to the diversity gains. In addition, based on the estimated UD's activity and channel, we propose a MUSIC-based multi-subarray collaborative localization (MSCLoc) algorithm to estimate the coordinates of each active UD. Finally, simulation results verify the superiority of our proposed algorithms. To the best of our knowledge, this is the first paper that considers the joint non-orthogonal multiple access and localization problem. Our contributions are summarized as follows.

\begin{itemize}
\item \textbf{Adopt a mmWave XL-MIMO array for uplink access and localization:}
The entire mmWave XL-MIMO array is split into multiple subarrays with ultra-wide adjacent subarray spacing to guarantee line-of-sight (LoS) link transmission to the desired UDs. Meanwhile, the multiple synchronized subarrays naturally serve as multiple anchors for accurate localization. Furthermore, the near-field SNS property is analyzed in the channel model and hybrid beamforming with phase shift network (PC-PSN) is adopted to avert the prohibitive hardware costs as well as the power consumption \cite{Heath_WC20, LiaoAW_Tcom19}.

\item \textbf{Propose the StrBOMP algorithm for efficient AUD and CE: } By exploiting the block sparsity of subarrays and the near-field SNS property of XL-MIMO channel, the proposed StrBOMP algorithm can significantly outperform the Oracle least squares (LS) algorithm with known active UDs, which is commonly adopted as the CE lower bound in the literature \cite{Qiao_TVT}. In addition, for improving performance, the structure of the XL-MIMO is utilized to refine the coarsely estimated AUD and CE results. Also, GMMV sensing matrices are designed to further achieve much better AUD and CE performance than the MMV sensing matrices.

\item \textbf{Develop the MSCLoc algorithm for localizing the active UDs: }After obtaining the users' activity and channel, we estimate the TDoAs between the multiple LoS links based on the MUSIC algorithm. Also, the AoAs of the LoS links from the active UDs to the related subarrays can be obtained. Then, the proposed MSCLoc algorithm collaboratively exploits the TDoAs and AoAs of multiple subarrays to acquire the coordinates of the active UDs with improved accuracy. Due to the large aperture of XL-MIMO and the large bandwidth of mmWave, simulation results show that the proposed MSCLoc algorithm can achieve centimeter-level indoor localization accuracy.
\end{itemize}
}

 {\it Notations}:
 Boldface lower and upper-case symbols denote column vectors and matrices, respectively.
 $(\cdot)^{\rm T}$, $(\cdot)^{\rm *}$, $(\cdot)^{\rm H}$, $(\cdot)^{\dagger}$, $\left| \cdot \right|$, $ \left\lceil \cdot \right\rceil$, $\otimes$, $\odot$, ${\text{det}}(\cdot)$, and $\mathbb{E}(\cdot)$ denote the transpose, conjugate, Hermitian transpose, Moore-Penrose pseudo-inverse, modulus, integer ceiling operators, Kronecker product, Hadamard product, determinant operator, and expectation operator, respectively. $\bm{0}_n$ ($\bm{0}_{m\! \times\! n}$) and $\bm{1}_n$ ($\bm{1}_{m\! \times\! n}$) denote the vectors (matrices) of size $n$ ($m\! \times\! n$) with all the elements being $0$ and $1$, respectively. $\bm{I}_n$ denotes the $n\! \times\! n$ identity matrix.  ${\| {\bm{a}} \|_p}$ and ${\| {\bm{A}} \|_F}$ are the ${\ell_p}$-norm of ${\bm{a}}$ and the Frobenius norm of ${\bm{A}}$, respectively. ${\text{diag}}(\bm{a})$ denotes diagonal matrix with the elements of $\bm{a}$ as its diagonal entries, and ${\text{Bdiag}}([\bm{A}_1,\! \cdots\! , \bm{A}_n])$ denotes the block diagonal matrix with $\bm{A}_1$, $\cdots$,$\bm{A}_n$ as its block diagonal entries. $[\bm{a}]_m$ ($[{\cal Q}]_m$) denotes the $m$th element of the vector $\bm{a}$ (the ordered set ${\cal Q}$). $[\bm{A}]_{m,n}$ denotes the the $m$th-row and the $n$th-column element of $\bm{A}$. $\bm{a}_{[{\cal Q}]}$ is the sub-vector containing the elements of $\bm{a}$ indexed by an ordered set ${\cal Q}$.
 $\bm{A}_{[{\cal Q},:]}$ ($\bm{A}_{[:,{\cal Q}]}$) denotes the sub-matrices containing the rows (columns) of $\bm{A}$ indexed by an ordered set ${\cal Q}$. ${\rm vec}({\bm A})$ stacks the columns of ${\bm A}$ on top of each another.  ${\text{supp}}\{\cdot\}$ denotes the set of non-zeros elements of its argument. $|\cdot|_c$ is the cardinality of a vector or a set. $[K]$ denotes the integer set $\{1, 2, ..., K \}$. ${\cal U}[a,b]$ denotes the uniform distribution over an interval $a$ to $b$. {\color{black}We denote $[{\cal K},\widetilde{\cal K}]=\Xi\{\cdot\}$, where $\Xi$ returns the unique values of its argument as an ordered set ${\cal K}$, and the ordered set $\widetilde{\cal K}$ is the number of times each unique element appears. Del$\{{\cal K},{\cal Q}\}$ returns the data in set ${\cal K}$ that is not in ${\cal Q}$. $[{\cal Q}]_{\cal K}$ denotes the elements in ${\cal Q}$ that are indexed by the set ${\cal K}$.}

\section{System Model}\label{S2}

\begin{figure*}[t]
\vspace{-9mm}
\begin{center}
 \includegraphics[width=1.2\columnwidth,keepaspectratio]{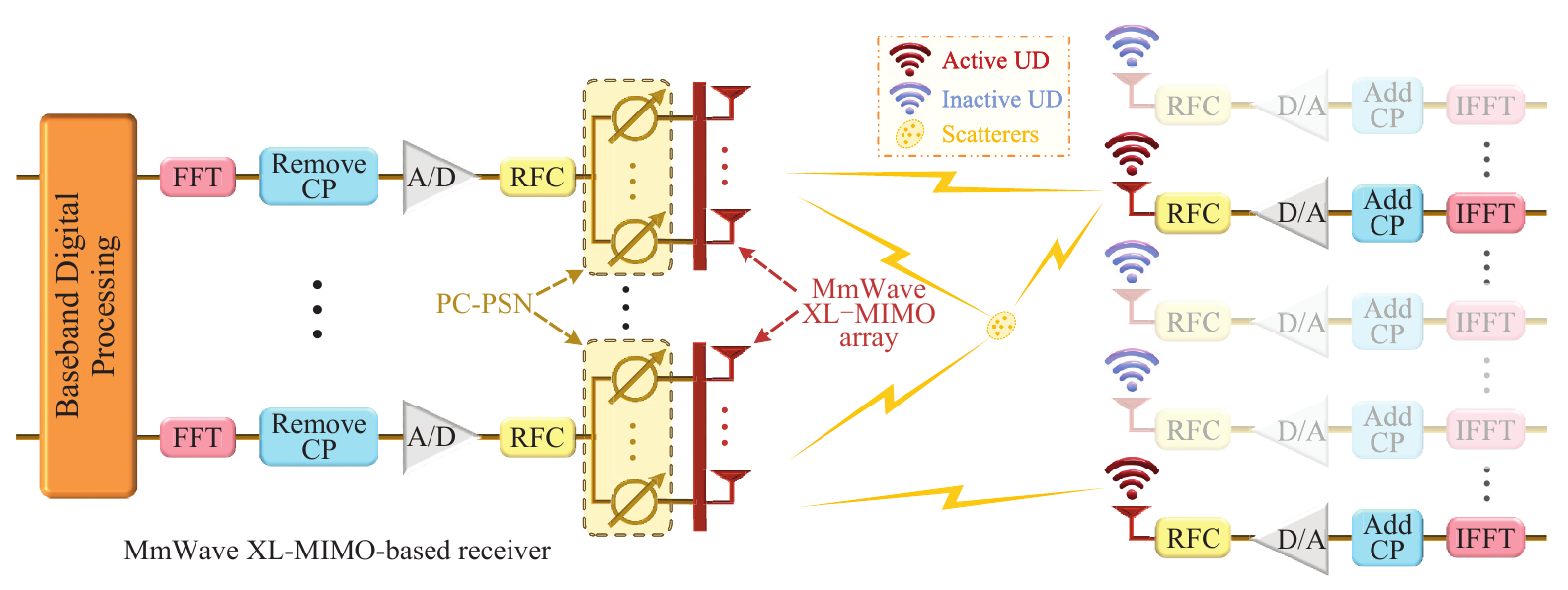}
\end{center}
\setlength{\abovecaptionskip}{-0.2mm}
 \captionsetup{font = {footnotesize}, name = {Fig.}, labelsep = period} 
\caption{The transceiver structure diagram for XL-MIMO-OFDM-based massive IoT access, where the XL-MIMO array equipped at the mmWave BS is partitioned into multiple subarrays and each subarray connects to its dedicated RFC through a PC-PSN.
The abbreviations FFT, IFFT, CP, D/A, and A/D represent fast Fourier transform, inverse FFT, cyclic prefix, digital-to-analog, and analog-to-digital, respectively.}
 \label{FIG2}
  \vspace{-6mm}
\end{figure*}

 In this section, we formulate the signal transmission model and the mmWave XL-MIMO channel model for massive IoT access in indoor scenario. The mmWave XL-MIMO array is embedded on the ceiling or the wall of the building, as shown in Fig.~\ref{FIG1}. To guarantee the existence of the LoS link, the mmWave XL-MIMO BS is partitioned into multiple subarrays with uniform linear array (ULA) structure for each subarray\footnote{Note that a simple yet feasible implementation of the mmWave XL-MIMO array adopted in this paper is the flexibly deployable radio stripes, which is a strip-shaped active massive MIMO array that can be attached on the surface of large construction elements \cite{ShaikZH_Tcom21,LopezOLA_TWC22}.}. Specifically, the XL-MIMO array deploys $N_{\rm BS}=M\times N_{\rm s}$ antennas and these antennas are equally divided into $M$ subarrays, where each subarray is allocated with one dedicated radio frequency chain (RFC) and $N_{\rm s}$ antennas, i.e., a partially connected structure.  Hence, the BS is equipped with $M$ RFCs, where each RFC connects to the corresponding subarray via a PC-PSN, as shown in Fig. \ref{FIG2}. The space between two adjacent subarrays is denoted by $\varDelta$ and the adjacent antenna spacing $d$ within the subarray is set to the half of the wavelength $\lambda$, i.e., $d\! =\! \lambda/2$. In addition, there are $K$ potential single-antenna UDs that can access the XL-MIMO BS under the grant-free random access protocol.

\subsection{Signal Model}\label{S2.1}

 Due to the key characteristic of the sporadic traffic in massive IoT access, only a small number,  $K_{\rm a}$, out of $K$ potential UDs (i.e., $K_{\rm a}\! \ll\! K$) are active within a certain coherence time slot \cite{KeML_TSP20}. To combat the impact of frequency selective fading caused by the multi-path delays in wideband mmWave networks, the OFDM technique is applied and $G$ consecutive OFDM symbols are adopted for uplink access as well as localization.  {\color{black}Besides, we consider the comb-type pilot structure \cite{MIMO-OFDM}, where $P$ subcarriers among the total number of subcarriers $N_{\rm c}$ are uniformly selected to transmit pilot signals.} As such, the signal vector ${\bm y}_p^{(g)}\! \in\! \mathbb{C}^{M}$, $g \in [G]$, received at the BS from the $K_{\rm a}$ active UDs on the $p$th, $p \in [P]$, pilot subcarrier of the $g$th OFDM symbol can be expressed as
\begin{align}\label{y_pg} 
 {\bm y}_p^{(g)} &= \big( {\bm W}_{\rm RF}^{(g)}{\bm W}_{{\rm BB},p}^{(g)} \big)^{\rm H} \sum\limits_{k = 1}^K {\zeta_k {\bm h}_{k,p}s_{k,p}^{(g)}} + {\bm n}_p^{(g)} \nonumber\\
 &= \big( {\bm W}_p^{(g)} \big)^{\rm H} {\bm H}_{p} {\bm s}_p^{(g)} + {\bm n}_p^{(g)} ,
\end{align}
 where ${\bm W}_p^{(g)}\! =\! {\bm W}_{\rm RF}^{(g)}{\bm W}_{{\rm BB},p}^{(g)}\! \in\! \mathbb{C}^{N_{\rm BS}\! \times\! M}$ denotes the hybrid combining matrix adopted at the BS cascaded by the analog and digital combining matrices, i.e., ${\bm W}_{\rm RF}^{(g)}\! \in\! \mathbb{C}^{N_{\rm BS}\! \times\! M}$ and ${\bm W}_{{\rm BB},p}^{(g)}\! \in\! \mathbb{C}^{M\! \times\! M}$, respectively,
 and $\zeta_k$ is a binary activity indicator that follows $\zeta_k\! =\! 1$ when the $k$th UD is active, otherwise $\zeta_k\! =\! 0$.
 Furthermore, ${\bm h}_{k,p}\! \in\! \mathbb{C}^{N_{\rm BS}}$ and ${\bm s}_p^{(g)}\! =\! \big[ s_{1,p}^{(g)}, s_{2,p}^{(g)},\! \cdots\! , s_{K,p}^{(g)} \big]^{\rm T}\! \in\! \mathbb{C}^{K}$ denote the XL-MIMO channel vector and the pilot signal vector, respectively. ${\bm n}_p^{(g)}\! =\! \big( {\bm W}_p^{(g)} \big)^{\rm H} {\bar{\bm n}}_p^{(g)}$ is the combined noise vector, where ${\bar{\bm n}}_p^{(g)}\! \sim\! {\cal CN}({\bm 0}_{N_{\rm BS}},\sigma_{\rm n}^2{\bm I}_{N_{\rm BS}})$ denotes the complex-valued additive white Gaussian noise (AWGN) with zero mean and covariance matrix $\sigma_{\rm n}^2{\bm I}_{N_{\rm BS}}$.
 Furthermore, ${\bm H}_{p}\! =\! \left[ \zeta_1{\bm h}_{1,p}, \zeta_2{\bm h}_{2,p},\! \cdots\! , \zeta_K{\bm h}_{K,p} \right]\! \in\! \mathbb{C}^{N_{\rm BS}\! \times\! K}$ in (\ref{y_pg}) denotes the aggregated channel matrix.
 We define the UDs activity indicator vector ${\bm \zeta}\! =\! \left[ \zeta_1, \zeta_2,\! \cdots\! , \zeta_K \right]^{\rm T}\! \in\! \{0,1\}^{K}$ with $\| {\bm \zeta} \|_0\! =\! K_{\rm a}$.

\subsection{XL-MIMO Channel Model}\label{S2.2}

Considering that there are $L_k$ multipath components (MPCs) for the $k$th UD, the uplink near-field SNS channel vector ${\bm h}_{k,p}\! \in\! \mathbb{C}^{N_{\rm BS}}$ on the $p$th subcarrier, $k\in[K]$, $p\in[P]$, can be expressed as
\begin{equation}\label{h_kp} 
 {\bm h}_{k,p} = \sqrt{\frac{\gamma_k}{\gamma_k + 1}} {\alpha_{k,1}} {\bm a}_{k,p,1} + \sqrt{\frac{1}{\gamma_k + 1}} \sum\limits_{l = 2}^{L_k} {\alpha_{k,l}}{{\bm a}_{k,p,l}} ,
\end{equation}
where {\color{black}$\alpha_{k,l}\! \sim\! {\cal CN}(0, 1)$ denotes the small scale fading coefficient for the $l$th MPC of the $k$th UD, $\forall l\in[L_k]$, $\gamma_k\geq 0$ is the Rician factor indicating the power ratio of small scale fading between the LoS path and the non-LoS (NLoS) MPCs \cite{MIMO-OFDM, RicianChannel}. ${{\bm a}_{k,p,l}}\! \in\! \mathbb{C}^{N_{\rm BS}}$ is the corresponding array response vector of the mmWave XL-MIMO array on the $p$th subcarrier. Note that the large scale fading is modeled into ${{\bm a}_{k,p,l}}$, which will be illustrated in detail in the following part.} 

\begin{figure*}[!tp]
\vspace{-9mm}
\begin{center}
\includegraphics[width=1.2\columnwidth,keepaspectratio]{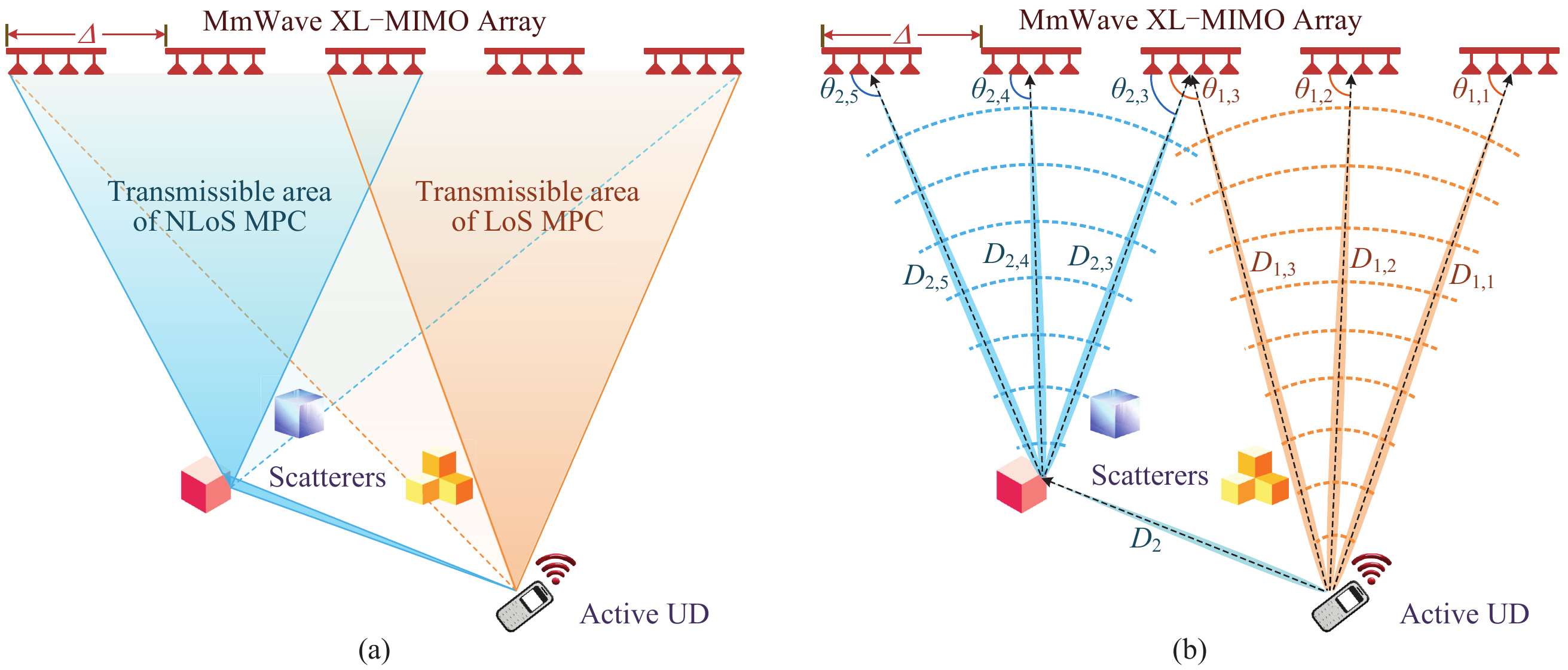}
\end{center}
\setlength{\abovecaptionskip}{-0.2mm}
\captionsetup{font = {footnotesize,color=black}, name = {Fig.}, labelsep = period} 
\caption{Schematic diagrams on the inherent characteristics of the mmWave XL-MIMO with multiple subarrays: (a) SNS property of the XL-MIMO channel; (b) Geometric channel modeling.}
\label{FIG3}
\vspace{-8mm}
\end{figure*}

 As shown in Fig.~\ref{FIG3}(a), influenced by ubiquitous obstacles and the extra-large array aperture, mmWave XL-MIMO channels exhibit the inherent near-field SNS property \cite{Heath_WC20}.
 In other words, the transmissible areas of LoS and/or NLoS MPCs from the same active UD may not cover the entire mmWave XL-MIMO array comprised of multiple subarrays.
 To capture this phenomenon, we define a binary near-field SNS vector ${\bm b}_{k,l}\! \in\! \mathbb{C}^{M}$ consisting of $0$ and $1$ that indicates whether the subarrays at the BSs can receive the electromagnetic (EM) signals from the $l$th MPC of the $k$th UD.
 Thus, the SNS array response vector ${{\bm a}_{k,p,l}}\! \in\! \mathbb{C}^{N_{\rm BS}}$ in (\ref{h_kp}) can be further written as
\begin{equation}\label{a_kpl} 
 {\bm a}_{k,p,l} = \left( {\bm b}_{k,l} \otimes {\bm 1}_{N_{\rm s}} \right) \odot \left[ {\bm c}_{k,p,l,1}^{\rm T},{\bm c}_{k,p,l,2}^{\rm T}, \cdots , {\bm c}_{k,p,l,M}^{\rm T} \right]^{\rm T} ,
\end{equation}
 where $l\in [L_k]$ and ${\bm c}_{k,p,l,m}\! \in\! \mathbb{C}^{N_{\rm s}}$ is the array response subvector associated with the $m$th subarray for $m\in[M]$. {\color{black}Recall to (\ref{h_kp}), considering all the MPCs together, we can denote the channel vector ${\bm h}_{k,p}\! \in\! \mathbb{C}^{N_{\rm BS}}$ as ${\bm h}_{k,p}\! =\! \left[ \pi_k^1({\bm h}_{k,p}^1)^{\rm T}, \pi_k^2({\bm h}_{k,p}^2)^{\rm T},\! \cdots\! , \pi_k^M({\bm h}_{k,p}^M)^{\rm T} \right]^{\rm T}$, where ${\bm h}_{k,p}^m\! \in\! \mathbb{C}^{N_{\rm s}}$ and $\pi_k^m\in\! \{0,1\}$ are the channel vector and the subarray activity indicator of the $m$-th subarray of the $k$-th UD on the $p$-th subcarrier, $m\in[M], \forall p,k$, respectively. Due to the SNS property, we have $\| {\bm \pi}_k \|_0\! \leq\! M$, $\forall k$, where ${\bm \pi_k}\! =\! \left[ \pi_k^1, \pi_k^2,\! \cdots\! , \pi_k^M \right]^{\rm T}\! \in\! \{0,1\}^{M}$ denotes the subarray activity indicator vector.}
 
 {\color{black}
 Furthermore, by analyzing the Rayleigh distance, we find that although some indoor UDs and scatterers are in the near-field region of the entire XL-MIMO array, they are in the far-field region of each subarray. Here, we take a specific example to justify this conclusion. Specifically, the Rayleigh distance can be calculated as $L_{{\rm Rd}}\! =\! 2A^2/\lambda$ \cite{Stutzman_Antenna12}, where $A$ denotes the array aperture.
 The system parameters are set to a carrier frequency $f_{\rm c}\! =\! 28$\,GHz, the number of antenna in a subarray
 $N_{\rm s}\! =\! 8$, adjacent subarray spacing $\varDelta\! =\! 5$ meter (m). 
 The apertures of a single subarray and an XL-MIMO array consisting of two subarrays are $A_1\! =\! N_{\rm s}\lambda/2\! \approx\! 0.04286$\,m and $A_2\! =\! \varDelta\! +\! 2{A_1}\! \approx\! 5.0857$\,m, respectively. 
 Their Rayleigh distances can be then calculated as $L_{{\rm Rd},1}\! =\! 2A_1^2/\lambda \! \approx\! 0.34286$\,m and $L_{{\rm Rd},2}\! =\! 2A_2^2/\lambda \! \approx\! 4828$\,m, respectively. Therefore, the subvector corresponding to each subarray, i.e., ${\bm c}_{k,p,l,m}$ in (\ref{a_kpl}), can be modelled by the far-field steering vectors, as shown Fig.~\ref{FIG3}(b). Note that due to the near-field assumption for the entire XL-MIMO array, the phase differences of different subarrays are modelled by the spherical EM waves.}
 
Specifically, ${\bm c}_{k,p,l,m}$ can be expressed in different ways for the LoS MPC and the NLoS MPCs. For the LoS MPC, i.e., $l\! =\! 1$, ${\bm c}_{k,p,1,m}\! \in\! \mathbb{C}^{N_{\rm s}}$ has the following form \cite{JinS_WCL20,DaiLL_CL21} as
\begin{align}\label{c_kplm_LoS} 
 {\bm c}_{k,p,1,m} \!\!=\!\!&\ \sqrt{\beta_{k,1,m}}e^{-{\textsf j}2\pi{\tau_{k,1,m}}\left( -\frac{f_{\rm s}}{2} + \frac{\left[ {(p-1){\bar P} + 1} \right] f_{\rm s}}{N_{\rm c}} \right)} \!\times\! {\bm e}\left(\theta_{k,1,m}\right),
\end{align}
 where $\bar P$ is the number of intervals between adjacent pilot subcarriers, $D_{k,1,m}$ is the transmission distance between the $m$th subarray and the $k$th UD, $\theta_{k,1,m}$ is the corresponding angle of arrival (AoA), as shown in Fig.~\ref{FIG3}(b). $\beta_{k,1,m}\! =\! \frac{G_{\rm s}\lambda^2}{\left(4\pi{D_{k,1,m}}\right)^2}$ and $\tau_{k,1,m}\! \sim\! {\cal U}(0,\tau_{\rm max}]$ are the large-scale fading coefficient
 and path delay, respectively, where $G_{\rm s}$ and $v_{\rm c}$ are the antenna gain and the speed of light. Moreover, assuming that the $m$th and $m'$th subarrays, $m,m'\in[M]$ and $m\neq m'$, have LoS links to the $k$th UD, we have $\tau_{k,1,m}-\tau_{k,1,m'}=\frac{D_{k,1,m}-D_{k,1,m'}}{v_{\rm c}}$. The steering vector related to the AoA $\theta_{k,1,m}$ is denoted by ${\bm e}\left(\theta_{k,1,m}\right)\! \in\! \mathbb{C}^{N_{\rm s}}$ and can be expressed as
\begin{equation}\label{e_klm} 
 {\bm e}({\theta_{k,1,m}})\! = \left[ 1,e^{-{\textsf j}\pi\cos\left(\theta_{k,1,m}\right)},\! \cdots\! ,e^{-{\textsf j}\left(N_{\rm s} - 1\right)\pi\cos\left(\theta_{k,1,m}\right)} \right]^{\rm T}.
\end{equation}

As for the NLoS MPCs, i.e., $l\! \in\! \{2,\! \cdots\! , L_k\}$, ${\bm c}_{k,p,l,m}\! \in\! \mathbb{C}^{N_{\rm s}}$ is formed as
\begin{align}\label{c_kplm_NLoS} 
 {\bm c}_{k,p,l,m} &= \sqrt{\beta_{k,l}^{(1)}}  
 \sqrt{\beta_{k,l,m}^{(2)}} e^{-{\textsf j}2\pi{\tau_{k,l,m}}\left( -\frac{f_{\rm s}}{2} + \frac{\left[ {(p-1){\bar P} + 1} \right] f_{\rm s}}{N_{\rm c}} \right)} \nonumber\\
 &~~~~~~~~~~~~~~~~~~~~~~~~~~~~~~~~~~~~\times{\bm e}\left(\theta_{k,l,m}\right) ,
\end{align}
 where $D_{k,l}^{(1)}$ and $D_{k,l,m}^{(2)}$ are the transmission distances from the $k$th UD to the $l$th scatterer and from this scatterer to the $m$th subarray, respectively,
 $\beta_{k,l}^{(1)}\! =\! \frac{\lambda^2}{\left(4\pi{D_{k,l}^{(1)}}\right)^2}$ and $\beta_{k,l,m}^{(2)}\! =\! \frac{G_{\rm s}\lambda^2}{\left(4\pi{D_{k,l,m}^{(2)}}\right)^2}$ are the corresponding large-scale fading coefficients,
 $\tau_{k,l,m} \! \sim\! {\cal U}(0,\tau_{\rm max}]$ is the total path delay from the $k$th UD to the $m$th subarray for the $l$th MPC. Note that the definitions of ${\bm e}\left(\theta_{k,l,m}\right)$ and $\theta_{k,l,m}$ are similar with those in (\ref{e_klm}).

\section{Proposed AUD, CE, and Localization Scheme with Near-Field XL-MIMO}\label{S3}
{\color{black}
In this section, we first formulate the joint AUD and CE problem into a CS problem. Then, we analyze the common subarray block sparsity of XL-MIMO channels and design the sensing matrices as well. Furthermore, the StrBOMP algorithm is proposed to efficiently estimate the active UDs and their channels. In addition, by adopting the outputs of StrBOMP algorithm and exploiting the XL-MIMO structure, we propose an MUSIC-based MSCLoc algorithm for estimating the locations of active UDs. Finally, the computational complexities of the proposed algorithms are analyzed.
}

\subsection{Problem Formulation}\label{S3.1}
To formulate the joint AUD and CE problem, we first vectorize the signal vector ${\bm y}_p^{(g)}\! \in\! \mathbb{C}^{M}$ in (\ref{y_pg}) on the $p$th pilot subcarrier of the $g$th OFDM symbol as
\begin{equation}\label{y_pg_vec} 
 {\bm y}_p^{(g)} = {\bm F}_p^{(g)}{{\bm h}_p} + {\bm n}_p^{(g)},~~p\in[P],
\end{equation}
 {\color{black}where ${\bm F}_p^{(g)}\! =\! \big( {\bm s}_p^{(g)} \big)^{\rm T}\! \otimes\! \big( {\bm W}_p^{(g)} \big)^{\rm H}\! \in\! \mathbb{C}^{M\! \times\! KN_{\rm BS}}$ and ${{\bm h}_p}\! =\! {\rm vec}\left( {\bm H}_{p} \right)\! =\! \left[ \zeta_1{\bm h}_{1,p}^{\rm T}, \zeta_2{\bm h}_{2,p}^{\rm T},\! \cdots\! , \zeta_K{\bm h}_{K,p}^{\rm T} \right]^{\rm T}\! \in\! \mathbb{C}^{KN_{\rm BS}}$. }
 
Taking the insufficient observations obtained by the limited RFCs of PC-PSN XL-MIMO architecture within one OFDM symbol into consideration, we aggregate the received signal matrices in $G$ OFDM symbols, i.e., ${\bm y}_p^{(g)}$ for $1\! \le\! g\! \le\! G$. Then, the stacked signal vector ${\bm y}_p\! =\! \big[ ({\bm y}_p^{(1)})^{\rm T}, ({\bm y}_p^{(2)})^{\rm T},\! \cdots\! , ({\bm y}_p^{(G)})^{\rm T} \big]^{\rm T}\! \in\! \mathbb{C}^{GM}$ is expressed by
\begin{equation}\label{y_p_a} 
 {\bm y}_p = {\bm F}_p{\bm h}_p + {\bm n}_p,~~p\in[P],
\end{equation}
 where ${\bm F}_p\! =\! \big[ ({\bm F}_p^{(1)})^{\rm T}, ({\bm F}_p^{(2)})^{\rm T},\! \cdots\! , ({\bm F}_p^{(G)})^{\rm T} \big]^{\rm T}\! \in\! \mathbb{C}^{GM\! \times\! KN_{\rm BS}}$
 and ${\bm n}_p$ are the sensing matrix associated with the $p$th pilot subcarrier and the corresponding noise vector, respectively. {\color{black}
 Note that the channel is commonly assumed invariant within the channel coherence time, hence we assume the channel vector ${\bm h}_p=\left[ \zeta_1{\bm h}_{1,p}^{\rm T}, \zeta_2{\bm h}_{2,p}^{\rm T},\! \cdots\! , \zeta_K{\bm h}_{K,p}^{\rm T} \right]^{\rm T}$, $\forall p\in[P]$, is constant within $G$ OFDM symbols. 
 The target of joint AUD and CE is to estimate the active UDs and the related channels, based on the sensing matrix ${\bm F}_p$ and the received signal ${\bm y}_p$, $p\in[P]$. Then, the localization target is to estimate the position of each active UD based on its estimated channel.
 However, since $GM\! \ll\! KN_{\rm BS}$, estimating the UDs activity indicator $\zeta_k$ (AUD) and reconstructing the channel vectors of active UDs (CE) from (\ref{y_p_a}), $p\in[P]$, is a challenging underdetermined problem.

\subsection{Common Subarray Block Sparsity of XL-MIMO Channels}\label{S3.2}

Fortunately, the number of active UDs typically accounts for a small fraction of all the potential UDs, i.e,  $\| {\bm \zeta} \|_0\! =\! K_{\rm a}\! \ll\! K$. {\color{black}
 In addition, according to (\ref{a_kpl}), due to the near-field SNS property of XL-MIMO channels, we have $\| {\bm \pi}_k \|_0\! \leq\! M$, i.e., only part of the subarrays are active for each active UD. Hence, the support sets of $\{ {{\bm h}_p} \}_{p=1}^P$ satisfies}
\begin{equation}\label{supp_hp_max} 
  \big| {\text{supp}}\{{{\bm h}_p}\} \big|_c \le K_{\rm a}N_{\rm s}M \ll KN_{\rm s}M=KN_{\rm BS},~~\forall p.
\end{equation}
{\color{black}Note that the channel vector relate to each subarray can be seen as a block with $N_{\rm s}$ elements. Hence, $\{ {{\bm h}_p} \}_{p=1}^P$, $\forall p\in[P]$, exhibits the subarray block sparsity.}

Furthermore, due to the same support set, ${\text{supp}}\{{\bm \zeta}\}$, shared by all the pilot subcarriers, ${{\bm h}_p}$, $p\in[P]$, exhibits the common sparsity in the frequency domain that is given by
\begin{align} 
 {\text{supp}}\{{{\bm h}_1}\} = {\text{supp}}\{{{\bm h}_2}\} = \cdots = {\text{supp}}\{{{\bm h}_P}\}  \label{supp_hp}.
\end{align}

{\color{black}Considering both (\ref{supp_hp_max}) and (\ref{supp_hp}), here we call the sparsity of the XL-MIMO channels as the {\it common subarray block sparsity}. Fig.~\ref{FIG4(a)} depicts the common sparsity of the channel matrix ${\bm H}\! =\! \left[ {{\bm{h}}_1}, {{\bm{h}}_2},\! \cdots\!, {{\bm{h}}_P} \right]\! \in\! \mathbb{C}^{KN_{\rm BS}\! \times\! P}$.
 It can be seen that the spatial-frequency domain channel vectors at all the pilot subcarriers share the common UDs sparse support sets. Furthermore, the channel matrix of the $4$th active UD is shown in Fig.~\ref{FIG4(b)}, where the subarray block sparsity and the near-field SNS property can be observed, i.e., the $9$th subarray is inactive. Hence, based on the common block sparsity of XL-MIMO channels, we can treat the joint AUD and CE problems in (\ref{y_p_a}) as the CS-based support detection and sparse signal recovery problems \cite{Tropp_SOMP06,GaoZ_TSP15}.}

\begin{figure}[!tp]
\centering
{\subfigure[]{ \label{FIG4(a)}
\begin{minipage}[t]{0.45\columnwidth}
\centering
\includegraphics[width=1\columnwidth]{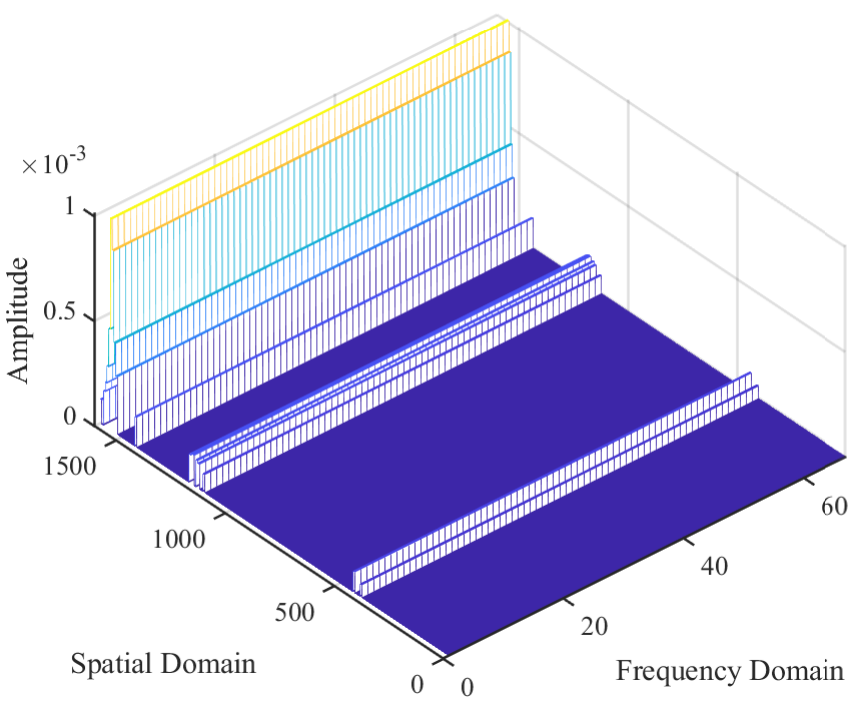}
\end{minipage}
}}
{\subfigure[]{ \label{FIG4(b)}
\begin{minipage}[t]{0.45\columnwidth}
\centering
\includegraphics[width=1\columnwidth]{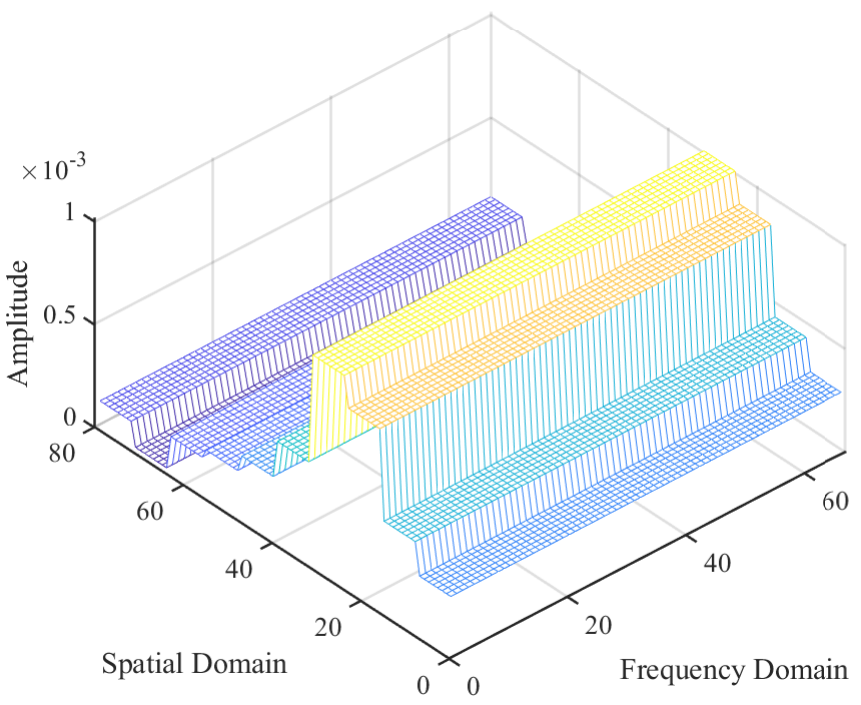}
\end{minipage}
}}
\setlength{\abovecaptionskip}{-0.2mm}
 \captionsetup{font = {footnotesize}, name = {Fig.}, labelsep = period} 
\caption{Common block sparsity of XL-MIMO channels at the spatial-frequency domain, where the system parameters are set to $M\! =\! 10$, $N_{\rm s}\! =\! 8$, $N_{\rm BS}\! =\!80$, $K\! =\! 20$, $K_{\rm a}\! =\! 4$, and $P\! =\! 67$: (a) Spatial-frequency domain channel matrix ${\bm H}$; (b) XL-MIMO channel associated with the $4$th active UD.}
\label{FIG4}
 \vspace{-6mm}
\end{figure}

\subsection{Sensing Matrix Design}\label{S3.3}

 To acquire the reliable reconstruction of sparse signals, the sensing matrices ${\bm F}_p^{(g)}\! =\! \big( {\bm s}_p^{(g)} \big)^{\rm T}\! \otimes\! \big( {\bm W}^{(g)} \big)^{\rm H}, p\in[P]$, $g\in[G]$, in (\ref{y_pg_vec}) should be designed elaborately to ensure the favorable restricted isometry property (RIP) \cite{GaoZ_TSP15}.
 Specifically, for the $p$th pilot subcarrier of the $g$th OFDM symbol, the pilot signal vector ${\bm s}_p^{(g)}$ transmitted from all the active UDs and the hybrid combining matrix ${\bm W}_p^{(g)}$ applied to the BS can be designed sequentially as follows.
 In particular, the $k$th entry of ${\bm s}_p^{(g)}$, denoted by $s_{k,p}^{(g)}$, can be set to $s_{k,p}^{(g)}\! =\! \sqrt{P_{\rm tx}}e^{{\textsf j}\phi_{k,p}^{(g)}}$, where $P_{\rm tx}$ is the transmit power for each UD and the phase $\phi_{k,p}^{(g)}$ follows the uniform distribution from $0$ to $2\pi$, i.e., $\phi_{k,p}^{(g)}\! \sim\! {\cal U}[0, 2\pi)$, for $k\in[K]$, $p\in[P]$, and $g\in[G]$.
 To design ${\bm W}_p^{(g)}\! =\! {\bm W}_{\rm RF}^{(g)}{\bm W}_{{\rm BB},p}^{(g)}$, on the one hand, due to the PC-PSN adopted by the XL-MIMO-based transceiver of the mmWave BS, the analog combining matrix has the block diagonal form, i.e., ${\bm W}_{\rm RF}^{(g)}\! =\! {\text{Bdiag}}\big( \big[ {\bm w}_{{\rm RF},1}^{(g)}, {\bm w}_{{\rm RF},2}^{(g)},\! \cdots\! , {\bm w}_{{\rm RF},M}^{(g)} \big] \big)$,
 where the $n$th entry of ${\bm w}_{{\rm RF},m}^{(g)}\! \in\! \mathbb{C}^{N_{\rm s}}$ can be set to $\big[ {\bm w}_{{\rm RF},m}^{(g)} \big]_n\! =\! \frac{1}{\sqrt{N_{\rm s}}} e^{{\textsf j}\varphi_{m,n}^{(g)}}$ with $\varphi_{m,n}^{(g)}\! \sim\! {\cal U}[0, 2\pi)$, for $n\in[N_{\rm s}]$, $m\in[M]$, and $g\in[G]$.
 Obviously, ${\bm W}_{\rm RF}^{(g)}$ is a semi-unitary matrix, i.e., $\big( {\bm W}_{\rm RF}^{(g)} \big)^{\rm H}{\bm W}_{\rm RF}^{(g)}\! =\! {\bm I}_{M}, \forall g$.
 On the other hand, the digital combining matrix can be considered as an identity matrix, i.e., ${\bm W}_{{\rm BB},p}^{(g)}\! =\! {\bm I}_{M}, \forall p,g$.
 Thus, the hybrid combining matrix satisfies ${\bm W}^{(g)}\! =\! {\bm W}_p^{(g)}\! =\! {\bm W}_{\rm RF}^{(g)}, \forall p$, i.e., ${\bm W}^{(g)}$ is also a semi-unitary matrix. Clearly, on the basis of the CS theory \cite{Tropp_SOMP06}, the designed ${\bm s}_p^{(g)}$ and ${\bm W}_p^{(g)}$
 can guarantee that the entries of sensing matrices ${\bm F}_p^{(g)}, \forall p,g$ obey the independent and identically distributed (i.i.d.) complex Gaussian distribution with zero mean and unit variance \cite{GaoZ_TSP15},
 that is, i.i.d. ${\cal CN}(0, 1)$, resulting in accurate support set detection and signal reconstruction. {\color{black}For GMMV setting, the pilot signals of each UD are different among pilot subcarriers, resulting in different sensing matrices ${\bm F}_p^{(g)}, p\in[P]$. This provides additional diversity gains than MMV setting, i.e., each UD adopts the same pilot signal among pilot subcarriers.}

{\color{black}Note that the combining matrix ${\bm W}^{(g)}$ is generated at the BS, while the pilot signals $s_{k,p}^{(g)}$, $\forall k,p,g$, is pre-defined and known at both the UDs and the BS. Hence, it is obvious that the BS can have access to the sensing matrices ${\bm F}_p^{(g)}\! =\! \big( {\bm s}_p^{(g)} \big)^{\rm T}\! \otimes\! \big( {\bm W}^{(g)} \big)^{\rm H}$. Here, we generate the pilot signals/random access preambles from complex Gaussian distribution as adopted in literature \cite{KeML_JSAC21,MYK,IimoriH_TWC22,Shim_Tcom19,YuanXJ_TWC20,YuW_TSP18_P1,
KeML_TSP20,YuanXJ_TWC22, ChenW_TWC22, DuY_WCL, WangY_IoTJ22, WuL_IoTJ21, Qiao_TVT,ShaoXD_TSP20,DjelouatH_WCL21, ChenXM_CL20, MaoZ_IoTJ22,Shim_IoTJ22,mmW-GF-gao2,mmW-GF-shao1}. According to \cite{Golay1,Golay2,Pengyu}, Golay sequences can provide at most 3\,dB peak-to-average power ratio, while achieving the similar AUD and CE performance to that of Gaussian random sequences. Hence, Golay sequences can be an alternative option for practical systems. }

\SetAlFnt{\footnotesize}
\SetAlCapFnt{\normalsize}
\SetAlCapNameFnt{\normalsize}
\begin{algorithm}[!t]
{\color{black}
\caption{Proposed StrBOMP Algorithm}\label{ALG2}
\LinesNumbered
\KwIn{Received signal vectors ${\bm y}_p\in\! \mathbb{C}^{M}$, sensing matrices ${\bm F}_p\in\! \mathbb{C}^{M\! \times\! KN_{\rm BS}}$, $ p\in[P]$, and pre-specified threshold $\epsilon_{\rm th}$}

\KwOut{The estimated UDs activity indicator vector $\widehat{\bm \zeta}$ and the estimated channel $\widehat{{\bm h}_p}\in\! \mathbb{C}^{KN_{\rm BS}}$, $ p\in[P]$}

 {\bf Initialization}: Iteration index $i\! =\! 1$, residual mean value $\epsilon^0\! =\! 2\epsilon_{\rm th}$, active block support set ${\cal S}^{0}\! =\! \emptyset$, non-zero elements set ${\cal I}^{0}\! =\! \emptyset$, activity indicator vector $\widehat{\bm \zeta}\! =\! {\bm 0}_{K}$, residual vectors ${\bm r}_p^{0}\! =\! {\bm y}_p$, and the channel vectors $\widehat{{\bm h}_p}\! =\! {\bm 0}_{KN_{\rm BS}}$, $p\in[P]$\;\label{A1_Init}
 
\While{$1$}
{
\label{TerCond}
${\bm d}\! =\! \sum\nolimits_{p=1}^P{ \big| {\bm F}_p^{\rm H}{\bm r}_p^{i - 1} \big| }$;  \%\ Calculate the correlations \label{blockCorr}

$s^{\star}\! =\! \big\{ s~\big{|}~\text{max}_{s\in[KM]}\sum\nolimits_{n=(s-1)N_{\rm s}+1}^{sN_{\rm s}}{ [{\bm d}]_n } \big\}$;  \%\ Block support estimation\label{SuppEst}

$\widetilde{\cal I}\! =\! \big\{ (s^{\star}\! -\! 1)N_{\rm s}\! +\! 1, (s^{\star}\! -\! 1)N_{\rm s}\! +\! 2,\! \cdots\! , s^{\star}N_{\rm s}\big\}$;  \%\ Indices of the elements within the block\label{set1}

${\cal S}^{i}\! =\! {\cal S}^{i-1} \cup s^{\star}$, and ${\cal I}^{i}\! =\! {\cal I}^{i-1} \cup \widetilde{\cal I}$;   \%\ Update the block support set and the set of non-zero elements\label{set2}

$\forall p\in[P]$:~~$\widehat{{\bm h}_p}_{[{\cal I}^{i}]}\! =\! {{\bm F}_p}_{[:,{\cal I}^{i}]}^{\dagger}{\bm y}_p$  \%\ Channel estimation via LS\label{LS}\;

$\forall p\in[P]$:~~${\bm r}_p^{i}\! =\! {\bm y}_p\! -\! {{\bm F}_p}_{[:,{\cal I}^{i}]}\widehat{{\bm h}_p}_{[{\cal I}^{i}]}$;  \%\  Residual update\label{ResUpdate}

$\epsilon^i\! =\! \frac{1}{GMP}\sum\nolimits_{p=1}^P{ \big({\bm r}_p^{i}\big)^{\rm H}{\bm r}_p^{i} }$  \%\ Calculate the residual mean value\;\label{meanCal}

$i\! =\! i\! +\! 1$\;

\If{$\epsilon^{i-1}\leq\epsilon_{\rm th}$ {\rm or} $\epsilon^{i-2}\leq\epsilon^{i-1}$}
{\label{StopC1}

$\{{\cal K},\widetilde{\cal K}\} =\Xi\{\lceil {{\cal S}^{i-1}/M} \rceil$\}, $\widehat{K_a}=\left| {\cal K} \right|_c$;  \%\ Obtain active UDs and number of active subarrays of each UD \label{ActSet}

${\cal J}\! =\! \big\{ [{\cal K}]_j~\big{|}~[\widetilde{\cal K}]_j=1,~j\in[\widehat{K_a}] \big\}$;  \%\ Select the active UDs that only have one related active subarray \label{SA1_set}

\If{$\left| {\cal J} \right|_c>1$}{\label{StopC2}

   ${\cal N}\! =\! \big\{ n~\big{|}~[\lceil {{\cal S}^{i-1}/M} \rceil]_n=j,~n\in[\left| {\cal S}^{i-1} \right|_c],~\forall j\in{\cal J} \big\}$ \; \label{delStart}
   
   $\widetilde{\cal I}\! =\! \big\{ ([{\cal S}^{i-1}]_n\! -\! 1)N_{\rm s}\! +\! 1, ([{\cal S}^{i-1}]_n\! -\! 1)N_{\rm s}\! +\! 2,\! \cdots\! , [{\cal S}^{i-1}]_nN_{\rm s}~\big{|}~n\in{\cal N}\big\}$\;

   ${\cal S}^{i}\! =\! \text{Del}\{ {\cal S}^{i-1}, [{\cal S}^{i-1}]_{\cal N}\}$, and ${\cal I}^{i}\! =\! \text{Del}\{{\cal I}^{i-1} , \widetilde{\cal I}\}$;  \%\ Delete the false alarm UDs \label{Del}

   \text{Run line \ref{LS} and \ref{ActSet}};  \%\ Refine the estimated channel vectors and active UDs set \label{Refine1}
  }
  \ElseIf{$\left| {\cal J} \right|_c=1$}{\label{StopC3}
  
    \text{Run line \ref{blockCorr}} \;\label{AddStart}
    
    $s^{\star}\! =\! \big\{ s~\big{|}~\mathop{\rm max}\limits_{s\in\{({\cal J}-1)M+1,...,{\cal J}M\}}\sum\nolimits_{n=(s-1)N_{\rm s}+1}^{sN_{\rm s}}{ [{\bm d}]_n } \big\}$;  \%\ Select the second active subarray of this UD \label{Sel_new}

    \text{Run line \ref{set1},~\ref{set2},~and \ref{LS}};  \%\ Refine the estimated channel vectors \label{Refine2}
  }
  \label{EndRefine}

$\widehat{\bm \zeta}_{[{\cal K}]}=1$;  \%\ Acquire the estimated UDs activity indicator vector \label{ActIndic}

{$\bf Break$};  \%\ Terminates the while-loop

}

}

\label{endFor}
{\bf Return}: The estimated activity indicator vector $\widehat{\bm \zeta}$ and the estimated channel $\widehat{{\bm h}_p}$, $ p\in[P]$ \label{A1_end}
}
\end{algorithm}

\subsection{Proposed StrBOMP Algorithm for Joint AUD and CE}\label{S3.5}

{\color{black}
The proposed StrBOMP algorithm is summarized in {\textbf{Algorithm~\ref{ALG2}}} for solving the sparse signal recovery problem in (\ref{y_p_a}). 
Specifically, according to the common block sparsity of mmWave XL-MIMO channels in different subcarriers, the correlations can be calculated in {\it{line}}~$\it \ref{blockCorr}$. Then, the possible block support with the maximum correlation value is selected in {\it{line}}~$\it \ref{SuppEst}$. To exploit the SNS property, i.e., $\| {\bm \pi}_k \|_0\! \leq\! M$, $\forall k$, each block corresponds to one subarray. In {\it{line}}~$\it \ref{set1}$, we obtain the indices of elements corresponding to the selected blocks. In {\it{line}}~$\it \ref{set2}$, we update the block support set ${\cal S}$ and the set of non-zero elements ${\cal I}$. Furthermore, elements with indices ${\cal I}$ in the channel vector ${\bm h}_p$, $\forall p\in[P]$, are estimated via LS in {\it{line}}~$\it \ref{LS}$. Then, the residual vectors are updated in {\it{line}}~$\it \ref{ResUpdate}$. In {\it{line}}~$\it \ref{meanCal}$, the residual mean value $\epsilon$ is calculated. If {\it{line}}~$\it \ref{StopC1}$ is satisfied, i.e., the residual mean value is smaller than the predefined threshold or larger than that in the last iteration, the algorithm gets out of the while loop. 

In {\it{line}}~$\it \ref{ActSet}$, based on the block support set ${\cal S}$, we acquire the estimated set of active UDs ${\cal K}$ and the number of active subarrays of each active UD. Note that the number of subarrays $M$ in XL-MIMO and the adjacent subarray spacing $\varDelta$ are quite large, hence the number of active subarrays of each active UD can be easily larger than one. In the following {\it{lines}}~$\it \ref{SA1_set}-\ref{EndRefine}$, we refine the active UDs set and the related channel vectors according to this prior information. Specifically, in {\it{line}}~$\it \ref{SA1_set}$, we obtain ${\cal J}$, which is the set of active UDs that only have one related active subarray. Then, if $\left| {\cal J} \right|_c>1$, it is likely that the active UDs in ${\cal J}$ are falsely alarmed UDs. Hence, we conduct {\it{lines}}~$\it \ref{delStart}-\ref{Refine1}$ to delete the false alarm UDs and refine the active UDs set as well as the estimated channel vectors. In addition, if $\left| {\cal J} \right|_c=1$, it is likely that the active subarrays of this active UD are miss detected. Hence, we perform {\it{lines}}~$\it \ref{AddStart}-\ref{Refine2}$ to add the second active subarray for this UD and refine the estimated channel vectors. In {\it{lines}}~$\it \ref{ActIndic}$, the estimated activity indicator vector $\widehat{\bm \zeta}$ is acquired based on ${\cal K}$. Finally, the outputs are $\widehat{\bm \zeta}$ and the estimated channel vectors $\widehat{{\bm h}_p}$, $ p\in[P]$. 
}

In {\it{line}}~$\it \ref{StopC1}$, the proposed StrBOMP algorithm adopts a pre-specified threshold $\epsilon_{\rm th}$ related to the noise variance as the stopping criterion, which can be exploited to adaptively acquire the active UDs set. Specifically, to determine such a threshold $\epsilon_{\rm th}$, we first analyze the estimation of the noise variance $\sigma_{\rm n}^2$. Denote ${\bm x}_p\! =\! {\bm F}_p{\bm h}_p$ as the signal vector, (\ref{y_p_a}) can be rewritten as ${\bm y}_p\! =\! {\bm x}_p\! +\! {\bm n}_p$. Based on the combined noise vector ${\bm n}_p^{(g)}\! =\! \big( {\bm W}_p^{(g)} \big)^{\rm H} {\bar{\bm n}}_p^{(g)}$ in (\ref{y_pg}) and the semi-unitary matrix ${\bm W}^{(g)}\! =\! {\bm W}_p^{(g)}, \forall p$, we have ${\bm n}_p\! =\! {\bar{\bm W}}^{\rm H} {\bar{\bm n}}_p$, 
 where ${\bar{\bm W}}\! =\! {\text{Bdiag}}\big( \big[ {\bm W}^{(1)}, {\bm W}^{(2)},\! \cdots\! , {\bm W}^{(G)} \big] \big)\! \in\! \mathbb{C}^{GN_{\rm BS}\! \times\! GM}$ and ${\bar{\bm n}}_p\! =\! \big[ ({\bar{\bm n}}_p^{(1)})^{\rm T}, ({\bar{\bm n}}_p^{(2)})^{\rm T},\! \cdots\! , ({\bar{\bm n}}_p^{(G)})^{\rm T} \big]^{\rm T}$ are the block diagonal combining matrix and the stacked AWGN vector, respectively.
 Due to the i.i.d. ${\bm n}_p$ for $p\in[P]$, we further consider the received signal vector ${\bm y}\! \in\! \mathbb{C}^{GMP}$ with stacked observations at all $P$ subcarriers, i.e., ${\bm y}\! =\! \big[ {\bm y}_1^{\rm T}, {\bm y}_2^{\rm T},\! \cdots\! , {\bm y}_P^{\rm T} \big]^{\rm T}\! =\! {\bm x}\! +\! {\bm n}$, where ${\bm x}\! =\! \big[ {\bm x}_1^{\rm T}, {\bm x}_2^{\rm T},\! \cdots\! , {\bm x}_P^{\rm T} \big]^{\rm T}$ and ${\bm n}\! =\! \big[ {\bm n}_1^{\rm T}, {\bm n}_2^{\rm T},\! \cdots\! , {\bm n}_P^{\rm T} \big]^{\rm T}\! =\! {\widetilde{\bm W}}^{\rm H} {\bar{\bm n}}$ with ${\widetilde{\bm W}}\! =\! {\bm I}_P\! \otimes\! {\bar{\bm W}}\! \in\! \mathbb{C}^{GN_{\rm BS}P\! \times\! GMP}$ and ${\bar{\bm n}}\! =\! \big[ {\bar{\bm n}}_1^{\rm T}, {\bar{\bm n}}_2^{\rm T},\! \cdots\! , {\bar{\bm n}}_P^{\rm T} \big]^{\rm T}$.
 Clearly, the stacked noise vector ${\bm n}$ still follows complex Gaussian distribution with the mean vector ${\bm 0}_{GMP}$ and the covariance matrix $\sigma_{\rm n}^2{\bm C}_{\rm n}$, i.e., ${\bm n}\! \sim\! {\cal CN}({\bm 0}_{GMP},\sigma_{\rm n}^2{\bm C}_{\rm n})$, where ${\bm C}_{\rm n}\! =\! {\widetilde{\bm W}}^{\rm H} {\widetilde{\bm W}}\! =\! {\bm I}_P\! \otimes\! \big( {\bar{\bm W}}^{\rm H} {\bar{\bm W}} \big)\! \in\! \mathbb{C}^{GMP\! \times\! GMP}$.
 Thus, considering the noise variance $\sigma_{\rm n}^2$ as a parameter to be estimated, the likelihood function of ${\bm y}$, denoted by ${\cal P}({\bm y};\sigma_{\rm n}^2)$, can be expressed as
{\color{black}
\begin{align} 
 &{\cal P}({\bm y};\sigma_{\rm n}^2)  \nonumber \\
 &= \frac{1}{{\pi}^{GMP}{\text{det}}\big( \sigma_{\rm n}^2 {\bm C}_n \big)} \exp \left( - ({\bm y} - {\bm x})^{\rm H} \big( {\sigma_{\rm n}^2 {\bm C}_{\rm n}} \big)^{-1} ({\bm y} - {\bm x}) \right) \nonumber \\
 &= \frac{\exp \left( - \frac{1}{\sigma_{\rm n}^2}({\bm y} - {\bm x})^{\rm H} \big(\big({\bm I}_P\big)^{-1}\otimes\big(  {\bar{\bm W}}^{\rm H} {\bar{\bm W}}\big)^{-1}\big) ({\bm y} - {\bm x}) \right)}{\big( {\pi} \sigma_{\rm n}^2 \big)^{GMP}{\text{det}}\big(  {\bm I}_P\! \otimes\! \big( {\bar{\bm W}}^{\rm H} {\bar{\bm W}} \big) \big)}  \nonumber \\
 &{\mathop=\limits^{(a)}} \frac{\exp \left( - \frac{1}{\sigma_{\rm n}^2} \sum\limits_{p=1}^P{ ({\bm y}_p - {\bm x}_p)^{\rm H} \big( {\bar{\bm W}}^{\rm H} {\bar{\bm W}} \big)^{-1} ({\bm y}_p - {\bm x}_p) } \right)}{\big( {\pi} \sigma_{\rm n}^2 \big)^{GMP}\big( {\text{det}}\big( {\bar{\bm W}}^{\rm H} {\bar{\bm W}} \big) \big)^P} 
 , \label{y_pdf}
\end{align}
}
where (a) is due to the property ${\text{det}}({\bm A}_{m\times m}\! \otimes\! {\bm B}_{n\times n})\! =\! \big( {\text{det}}({\bm A}) \big)^n \big( {\text{det}}({\bm B}) \big)^m$ \cite{Golub_Matrix13}.
 The log-likelihood function of ${\bm y}$, i.e., ${\cal L}({\bm y};\sigma_{\rm n}^2)\! =\! \ln{{\cal P}({\bm y};\sigma_{\rm n}^2)}$, can be further written as
\begin{align} 
 &{\cal L}({\bm y};\sigma_{\rm n}^2) = -GMP\ln{\big( \pi \sigma_{\rm n}^2 \big)} - P\ln{\big( {\text{det}}\big( {\bar{\bm W}}^{\rm H} {\bar{\bm W}} \big) \big)}  \nonumber\\
 &~~~~~~~~~~~~~~~~~~- \frac{1}{\sigma_{\rm n}^2} \sum\limits_{p=1}^P{ ({\bm y}_p - {\bm x}_p)^{\rm H} \big( {\bar{\bm W}}^{\rm H} {\bar{\bm W}} \big)^{-1} ({\bm y}_p - {\bm x}_p) } . \label{y_ln_pdf}
\end{align}
 Based on the ML criterion, the ML estimator of the noise variance, denoted by $\widehat\sigma_{\rm n}^2$, can thus be obtained by solving the log-likelihood equation $\partial{{\cal L}({\bm y};\sigma_{\rm n}^2)} \big/ \partial{\sigma_{\rm n}^2}\! =\! 0$ such that we have
\begin{align}\label{sigma_hat} 
 \widehat\sigma_{\rm n}^2 &= \frac{1}{GMP}\sum\limits_{p=1}^P{ 
 ({\bm y}_p - {\bm x}_p)^{\rm H} \big( {\bar{\bm W}}^{\rm H} {\bar{\bm W}} \big)^{-1} ({\bm y}_p - {\bm x}_p) } \nonumber\\
 &{\mathop=\limits^{(b)}} \frac{1}{GMP}\sum\limits_{p=1}^P{
 ({\bm y}_p - {\bm x}_p)^{\rm H} ({\bm y}_p - {\bm x}_p) },
\end{align}
 where the block diagonal combining matrix ${\bar{\bm W}}$ consisting of $\{ {\bm W}^{(g)} \}_{g=1}^{G}$ satisfies ${\bar{\bm W}}^{\rm H}{\bar{\bm W}}\! =\! {\bm I}_{GM}$ in equation (b) due to the hybrid combining matrices ${\bm W}^{(g)}, \forall g$, designed in Section~\ref{S3.3} being the semi-unitary matrices.

Furthermore, we relate the noise variance estimation $\widehat\sigma_{\rm n}^2$ in (\ref{sigma_hat}) with the pre-specified threshold $\epsilon_{\rm th}$ in Algorithm~\ref{ALG2}.
 For the processing of the $p$th subcarrier in the $i$th iteration, according to the LS estimation of block channel vector, i.e., ${{\bm h}}_{{\rm b},p}^i\! =\! {{\bm F}_p}_{[:,{\cal I}^{i}]}^{\dagger}{\bm y}_p$ in {\it{line}}~$\it 13$, the estimation of signal vector ${\bm x}_p$ at the $i$th iteration, denoted by ${\widehat{\bm x}_p^{i}}$, is given by 
\begin{equation}\label{y_p_hat} 
 {\widehat{\bm x}_p^{i}} = {{\bm F}_p}_{[:,{\cal I}^{i}]}{\widetilde{\bm h}}_{{\rm b},p}^i
 = {{\bm F}_p}_{[:,{\cal I}^{i}]} {{\bm F}_p}_{[:,{\cal I}^{i}]}^{\dagger} {\bm y}_p 
 = {\bm P}_p^i {\bm y}_p ,
\end{equation}
 where ${\bm P}_p^i\! =\! {{\bm F}_p}_{[:,{\cal I}^{i}]} {{\bm F}_p}_{[:,{\cal I}^{i}]}^{\dagger}\! \in\! \mathbb{C}^{GM\! \times\! GM}$ is the projection matrix of the selected sensing matrix ${{\bm F}_p}_{[:,{\cal I}^{i}]}$.
 Thus, the estimation error vector ${\bm r}_p^{i}$ can be expressed as
\begin{equation}\label{r_p_i} 
 {\bm r}_p^{i} = {\bm y}_p - {\widehat{\bm x}_p^{i}} 
 = \big( {\bm I}_{GM} - {\bm P}_p^i \big) {\bm y}_p
 = \big( {\bm P}_p^i \big)^\bot {\bm y}_p ,
\end{equation}
 where $\big( {\bm P}_p^i \big)^\bot\! =\! {\bm I}_{GM} - {\bm P}_p^i$ is the orthogonal projection matrix of ${\bm P}_p^i$.
 The physical interpretations of (\ref{y_p_hat}) and (\ref{r_p_i}) are that the estimated signal vector ${\widehat{\bm x}_p^{i}}$ and the estimation error vector ${\bm r}_p^{i}$ are the projection and orthogonal projection of the received signal vector ${\bm y}_p$ onto the subspace spanned by the selected sensing matrix ${{\bm F}_p}_{[:,{\cal I}^{i}]}$ \cite{Golub_Matrix13,Tropp_ICASSP06}, respectively.
 Therefore, the joint AUD and CE process of the proposed StrBOMP algorithm is to continuously find the orthogonal projection of the signal vector onto the subspace spanned by the iteratively selected sensing matrix,
 during which the active subarray support set and the channel vectors to be estimated can be acquired.
 After a sufficient number of iterations, i.e., $i\! =\! I$, the orthogonal projection of ${\bm y}_p$ onto the subspace spanned by ${{\bm F}_p}_{[:,{\cal I}^{I}]}$, denoted by ${\bm r}_p^{I}$, will no longer contain useful signals from the active UDs but only noise.
 Provided that the obtained active user support set and channel vectors are accurate enough, we can approximate the estimated signal vector ${\widehat{\bm x}}_p^{I}$ to the actual signal vector ${\bm x}_p$ \cite{Tropp_ICASSP06}, given by
\begin{equation}\label{x_p_approx} 
 {\bm x}_p \approx {\widehat{\bm x}}_p^{I}
 = {{\bm F}_p}_{[:,{\cal I}^{I}]} {{\bm h}}_{{\rm b},p}^{I} .
\end{equation}
 Substituting (\ref{x_p_approx}) into (\ref{sigma_hat}), we can obtain
\begin{align}\label{sigma_hat_approx} 
 \widehat\sigma_{\rm n}^2 &\approx \frac{1}{GMP}\sum\limits_{p=1}^P{ 
 \big( {\bm y}_p\! -\! {{\bm F}_p}_{[:,{\cal I}^{I}]} {{\bm h}}_{{\rm b},p}^{I} \big)^{\rm H} \big( {\bm y}_p\! -\! {{\bm F}_p}_{[:,{\cal I}^{I}]} {{\bm h}}_{{\rm b},p}^{I} \big) } \nonumber\\
 &= \frac{1}{GMP}\sum\limits_{p=1}^P{ 
 \big( {\bm r}_p^{I} \big)^{\rm H} {\bm r}_p^{I} }.
\end{align}

 Based on the aforementioned analysis, the ultimate residual mean value $\epsilon\! =\! \frac{1}{GMP}\sum\nolimits_{p=1}^P{ \big({\bm r}_p^{i}\big)^{\rm H}{\bm r}_p^{i} }$ for $i\! =\! I$ in {\it{line}}~$\it \ref{meanCal}$ of Algorithm~\ref{ALG2} is expected to approximate the estimated noise variance $\widehat\sigma_{\rm n}^2$.
 From the ML perspective, it is optimal to treat the actual noise variance as the termination threshold \cite{Tropp_SOMP06}, we can thus set the actual noise variance as the predefined threshold, i.e., $\epsilon_{\rm th}\! =\! \sigma_{\rm n}^2$.
 More intuitively, when the active subarray support set and channel vectors are estimated with sufficient accuracy, the useful signal components in the residual vectors will be eliminated except for the noise.
 Hence, the power of the residual noise (that is, the noise variance) can be approximated as the average of all the residual entries.

\subsection{Proposed MSCLoc Algorithm for Active UDs' Localization}\label{S3}
In this subsection, we design the localization algorithm for active UDs based on the outputs of Algorithm 1, i.e., estimated activity indicator vector $\widehat{\bm \zeta}$ and channel vectors $\widehat{{\bm h}_p}$, $ p\in[P]$, and the MUSIC algorithm \cite{Shao_sensing}. For simplicity, we will introduce the localization of any one of the detected active UDs as follows. Finally, the proposed MSCLoc algorithm is summarized in Algorithm \ref{ALG3}.

\subsubsection{AoAs and TDoAs Estimation} 
From Algorithm 1, suppose the $k$th ($k\in[K]$) UD is detected as active, where its estimated channel is denoted as $\widehat{\bm H}_k=[\widehat{\bm H}_{k,1}^{\rm T},\widehat{\bm H}_{k,2}^{\rm T},...,\widehat{\bm H}_{k,M}^{\rm T}]^{\rm T}\in\mathbb{C}^{N_{\rm BS}\times P}$ and $\widehat{\bm H}_{k,m}\in\mathbb{C}^{N_{\rm s}\times P}$ ($m\in[M]$) corresponds to the $m$th subarray. Then, for this UD, use vector ${\bm p}_k=[\|\widehat{\bm H}_{k,1} \|_F, \|\widehat{\bm H}_{k,2} \|_F,...,\|\widehat{\bm H}_{k,M} \|_F]^{\rm T}\in\mathbb{R}^{M}$ to denote the energy of each subarray's estimated channel. According to (\ref{a_kpl}), only some of the subarrays have LoS links to the active UDs due to the near-field SNS property. Based on ${\bm p}_k$, we can distinguish the subarrays that have LoS links from those that only have NLoS links. First, we normalize ${\bm p}_k$ by using min-max normalization \cite{TWCqiao}, then the subarrays that have LoS links can be determined if their normalized energy is larger than a predefined threshold $0<\phi<1$. Hence, we obtain the set of subarrays that have LoS links to the $k$th UD as
\begin{equation}\label{Set_LoS} 
\Omega_k=\left\{~m~\big{|}~\dfrac{[{\bm p}_k]_m-min({\bm p}_k)}{max({\bm p}_k)-min({\bm p}_k)}>\phi, m\in[M] \right\},
\end{equation}
where $max(\cdot)$ and $min(\cdot)$ return the maximum and minimum values of their arguments, respectively.

Furthermore, given any $m\in\Omega_k$, the corresponding estimated channel matrix $\widehat{\bm H}_{k,m}\in\mathbb{C}^{N_{\rm s}\times P}$ mainly consists of two parts, i.e., the LoS component and other components, which can be expressed as follows
\begin{align}\label{Hkm_rewrite} 
 \widehat{\bm H}_{k,m}&=\sqrt{\frac{\gamma_k\beta_{k,1,m}}{\gamma_k + 1}} {\alpha_{k,1}}{\bm e}\left(\theta_{k,1,m}\right)\times{\bm g}^{\rm T}\left(\tau_{k,1,m}\right)+{\bm Z}_{k,m} \nonumber\\
 &{\mathop=\limits^{\Delta}}~ {\bm H}_{k,m}^{\rm LoS}+{\bm Z}_{k,m},
\end{align}
where $\gamma_k$ and $\alpha_{k,1}\! \sim\! {\cal CN}(0, 1)$ are the Rician factor and small-scale fading coefficient defined in (\ref{h_kp}), $\beta_{k,1,m}$ and ${\bm e}\left(\theta_{k,1,m}\right)\in\mathbb{C}^{N_{\rm s}}$ are the large-scale fading coefficient and the steering vector defined in (\ref{c_kplm_LoS}), ${\bm Z}_{k,m}\in\mathbb{C}^{N_{\rm s}\times P}$ consists of the NLoS components and the channel estimation error. Moreover, ${\bm g}\left(\tau_{k,1,m}\right)\in\mathbb{C}^{P}$ denotes the phase shifts in different pilot subcarriers. According to (\ref{c_kplm_LoS}), ${\bm g}\left(\tau_{k,1,m}\right)\in\mathbb{C}^{P}$ can be expressed as
\begin{align}\label{ggg}
    {\bm g}(\tau_{k,1,m})\! &= \!\!\big[e^{-{\textsf j}2\pi{\tau_{k,1,m}}\psi(1)},e^{-{\textsf j}2\pi{\tau_{k,1,m}}\psi(2)},\! \cdots\! ,\nonumber \\
 &~~~~~~~~~~~~~~~~~~~~~~~~~~~~~~~e^{-{\textsf j}2\pi{\tau_{k,1,m}}\psi(P)} \big]^{\rm T},
\end{align} 
where $\psi(p)=-\frac{f_{\rm s}}{2} + \frac{[ {(p-1){\bar P} + 1} ] f_{\rm s}}{N_{\rm c}}$, $p\in[P]$, $\tau_{k,1,m}\! \sim\! {\cal U}(0,\tau_{\rm max}]$ is the delay of the related LoS link defined in (\ref{c_kplm_LoS}). 

By performing the MUSIC algorithm, we can obtain the delay $\tau_{k,1,m} $ and AoA $\theta_{k,1,m}$, $m\in\Omega_k$, from the estimated channel matrix $\widehat{\bm H}_{k,m}$. Specifically, the covariance matrices of $\widehat{\bm H}_{k,m}$ is given by
\begin{align}\label{Cov} 
 &{\bm R}_{\widehat{\bm H}_{k,m}}\! = \dfrac{1}{P}\widehat{\bm H}_{k,m}\widehat{\bm H}_{k,m}^{\rm H}\nonumber\\
 &=\dfrac{\gamma_k\beta_{k,1,m}\left| \alpha_{k,1} \right|}{P(\gamma_k + 1)}{\bm e}(\theta_{k,1,m}){\bm g}^{\rm T}(\tau_{k,1,m}){\bm g}^{\rm *}(\tau_{k,1,m}){\bm e}^{\rm H}(\theta_{k,1,m}) \nonumber\\
 &~~~~~~~~~~~~~~~~~~~~~~~~~~~~~~~~~~~~~~~~~~~~~~~~~~~~~~~~~+ {\bm R}_{{\bm Z}_{k,m}},
\end{align}
where ${\bm R}_{{\bm Z}_{k,m}}=\dfrac{1}{P}{\bm Z}_{k,m}{\bm Z}_{k,m}^{\rm H}$. Then, performing eigenvalue decomposition (EVD) on ${\bm R}_{\widehat{\bm H}_{k,m}}$, yields
\begin{equation}\label{EVD} 
 {\bm R}_{\widehat{\bm H}_{k,m}}\! = 
\begin{bmatrix}
{\bm U}_s&{\bm U}_z
\end{bmatrix}
\begin{bmatrix}
{\bm \Sigma}_s &    \\
	& {\bm \Sigma}_z
\end{bmatrix}
\begin{bmatrix}
{\bm U}_s^{\rm H}   \\
{\bm U}_z^{\rm H}
\end{bmatrix}
,
\end{equation}
where ${\bm U}_s\in\mathbb{C}^{N_{\rm s}}$ and ${\bm U}_z\in\mathbb{C}^{N_{\rm s}\times (N_{\rm s}-1)}$ are the eigenvectors that span the signal and noise subspace, respectively. As ${\bm e}(\theta_{k,1,m})$ is orthogonal to ${\bm U}_z$, the AoA $\theta_{k,1,m}$ is estimated by MUSIC algorithm as
\begin{equation}\label{OptTheta} 
 \widehat{\theta}_{k,1,m}=\left\{ \theta~\big{|}~{\rm max}_{\theta\in(0,\pi]}\dfrac{1}{{\bm e}^{\rm H}(\theta){\bm U}_z{\bm U}_z^{\rm H}{\bm e}(\theta)}\right\},
\end{equation}
where the commonly adopted hierarchical search method is employed to reduce the computational complexity \cite{Shao_sensing}.

As for the estimation of the TDoAs, we conduct EVD on covariance matrix ${\bm R}_{\widehat{\bm H}^{\rm H}_{k,m}}=\dfrac{1}{N_{\rm s}}\widehat{\bm H}_{k,m}^{\rm H}\widehat{\bm H}_{k,m}$ and obtain the related eigenvectors of noise subspaces, denoted as $\overline{\bm U}_z$. Hence, the path delay $\tau_{k,1,m}$ is estimated by the MUSIC algorithm as
\begin{equation}\label{OptDelay} 
 \widehat{\tau}_{k,1,m}=\left\{ \tau~\big{|}~{\rm max}_{\tau\in(0,\tau_{\rm max}]}\dfrac{1}{{\bm g}^{\rm T}(\tau)\overline{\bm U}_z\overline{\bm U}_z^{\rm H}{\bm g}^{\rm *}(\tau)}\right\}.
\end{equation}
Then, we can calculate the TDoA between the $m$th and the $m'$th subarrays, $m,m'\in\Omega_k$ and $m\neq m'$, which is expressed as $\widehat{\tau}_{k,1,m}-\widehat{\tau}_{k,1,m'}$. According to (\ref{c_kplm_LoS}), $\widehat{\tau}_{k,1,m}-\widehat{\tau}_{k,1,m'}=\frac{D_{k,1,m}-D_{k,1,m'}}{v_{\rm c}}$.

 \begin{figure}[!tp]
\vspace{-8mm}
\begin{center}
 \includegraphics[width=0.8\columnwidth,keepaspectratio]{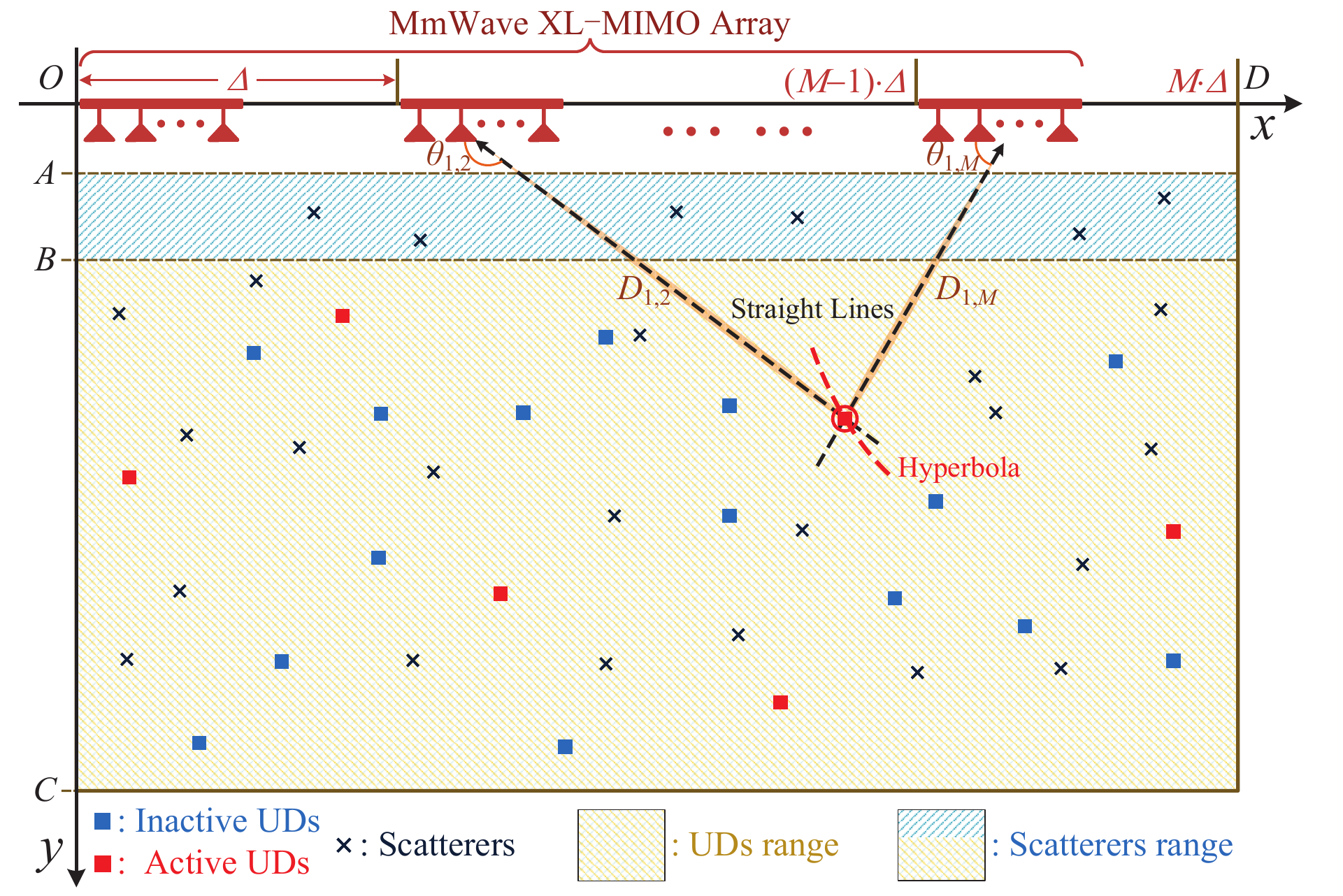}
\end{center}
\setlength{\abovecaptionskip}{-0.2mm}
 \captionsetup{font = {footnotesize}, name = {Fig.}, labelsep = period} 
\caption{Schematic diagram of the considered XL-MIMO-based indoor massive access scenario and the localization method for active UDs.}
 \label{FIG5}
 \vspace{-6mm}
\end{figure}

\subsubsection{Coordinates Estimation}
As shown in Fig.~\ref{FIG5}, without loss of generality, we consider a two-dimensional indoor space under Cartesian coordinate with the size of $\overline{OC}\! \times\! \overline{OD}$. The mmWave XL-MIMO array embedded on the top or on the wall has $M$ subarrays with ultra-wide adjacent subarray spacing $\varDelta$. We denote the coordinate of the center of the $m$th subarray as $(x_m,0)$, $m\in[M]$. The UDs are randomly distributed within the regions of $\{ (x,y)\vert 0\! \le\! x \! \le\! \overline{OD} \cup \overline{OB}\! \le\! y \! \le\! \overline{OC} \}$, while the scatterers are randomly located within the regions of $\{ (x,y)\vert 0\! \le\! x \! \le\! \overline{OD} \cup \overline{OA}\! \le\! y \! \le\! \overline{OC} \}$. The target is to estimate the coordinates of the active UDs based on the related AoAs and TDoAs.

Suppose that the coordinates of the $k$th UD is $(x_k,y_k)$. Based on the estimated AoAs $\widehat{\theta}_{k,1,m}$, $m\in\Omega_k$, we can obtain $|\Omega_k|_c$ linear equations as follows
\begin{equation}\label{Lines} 
x_k+\cot(\widehat{\theta}_{k,1,m}) y_k=x_m,~m\in \Omega_k,
\end{equation}
which means that there are $|\Omega_k|_c$ straight lines passing through the $k$th UD.

Furthermore, based on the estimated TDoAs, we can obtain another $|\Omega_k|_c-1$ hyperbolic equations that also pass through the $k$th UD, which can be expressed as
\begin{align}\label{hyperbolas} 
y_k\left(\dfrac{1}{\sin(\widehat{\theta}_{k,1,m})}- \dfrac{1}{\sin(\widehat{\theta}_{k,1,m_0})}\right)&=v_c\left(\widehat{\tau}_{k,1,m}-\widehat{\tau}_{k,1,m_0}\right),
\end{align}
where $m\in \Omega_k~ {\rm and}~ m\neq m_0$, $m_0$ is selected from $\Omega_k$. Note that the number of subarrays $M$ in XL-MIMO and the adjacent subarray spacing $\varDelta$ are quite large, hence $|\Omega_k|_c>1$ can be easily satisfied.

In addition, equations (\ref{Lines}) and (\ref{hyperbolas}) can be expressed in a compact form as
\begin{align}\label{LocMatrix} 
&\begin{bmatrix}
x_k \!&\! y_k
\end{bmatrix}
\underbrace{\begin{bmatrix}
0 \!\!& \!\!\cdots \!\!&\!\! 0 \!\!&\!\! 1 &\!\!\cdots\!\! &\!\! 1\\
\delta_1 \!\!& \!\!\cdots \!\!&\!\! \delta_{M'-1} \!\!&\!\! \cot(\widehat{\theta}_{k,1,[\Omega_k]_1})\!\! & \!\!\cdots\!\! &\!\! \cot(\widehat{\theta}_{k,1,[\Omega_k]_{M'}})
\end{bmatrix}
}_{{\bm \Phi}^{\rm T}_k}\nonumber\\
&~~~~~~~~~~~~~~~=
\underbrace{\begin{bmatrix}
\overline{d}_1 & \!\!\cdots\!\! & \overline{d}_{M'-1} & x_{[\Omega_k]_1} & \!\!\cdots\!\! & x_{[\Omega_k]_{M'}}
\end{bmatrix}
}_{{\bm \psi}^{\rm T}_k}
,
\end{align}
where $M'=|\Omega_k|_c$. Furthermore, as for $m'\in[M'-1]$, we have $\delta_{m'}=\frac{1}{\sin(\widehat{\theta}_{k,1,[\overline{\Omega}_k]_{m'}})}- \frac{1}{\sin(\widehat{\theta}_{k,1,m_0})}$ and $\overline{d}_{m'}=v_c(\widehat{\tau}_{k,1,[\overline{\Omega}_k]_{m'}}-\widehat{\tau}_{k,1,m_0})$, where $\overline{\Omega}_k=\{i~|~i\in\Omega_k, i\neq m_0\}$.
 
By exploiting the weighted LS algorithm, we can acquire the estimated coordinates of the $k$th UD as
 \begin{equation}\label{CoordEst} 
\begin{bmatrix}
\widehat{x}_k & \widehat{y}_k
\end{bmatrix}^{\rm T}
=\left({{\bm \Phi}^{\rm H}_k} {\bm Q}{{\bm \Phi}_k}\right)^{-1}{{\bm \Phi}^{\rm H}_k} {\bm Q} {{\bm \psi}_k},
\end{equation}
where ${\bm Q}=\text{diag}([\frac{1}{v_c\widehat{\tau}_{k,1,[\overline{\Omega}_k]_{1}}},\cdots,\frac{1}{v_c\widehat{\tau}_{k,1,[\overline{\Omega}_k]_{M'-1}}},\frac{1}{v_c\widehat{\tau}_{k,1,[{\Omega}_k]_{1}}},$ $\cdots,\frac{1}{v_c\widehat{\tau}_{k,1,[{\Omega}_k]_{M'}}}]^{\rm T})$ denotes the weights of different equations. It can be seen that the equation relates to a shorter transmission distance has a larger weight, which can further improve the estimation accuracy of the coordinates.

\SetAlFnt{\footnotesize}
\SetAlCapFnt{\normalsize}
\SetAlCapNameFnt{\normalsize}
\begin{algorithm}[!t]
\caption{Proposed MSCLoc Algorithm}\label{ALG3}
\LinesNumbered
\KwIn{The activity indicator vector $\widehat{\bm \zeta}$ and channel vectors $\widehat{{\bm h}_p}$, $ p\in[P]$, acquired from Algorithm \ref{ALG2}, the threshold $\phi$}
\KwOut{Estimated coordinates of active UDs $(\widehat{x}_k,\widehat{y}_k)$, $k\in\widehat{\cal S}$}
\For {$k\in\widehat{\cal S}$}{
$\widehat{\bm H}_k=[\widehat{\bm H}_{k,1}^{\rm T},\widehat{\bm H}_{k,2}^{\rm T},...,\widehat{\bm H}_{k,M}^{\rm T}]^{\rm T}={\widehat{\bm H}}_{[{\cal I}_k,:]}$, where ${\cal I}_k\! =\! \{ (k\! -\! 1)N_{\rm BS}\! +\! 1, (k\! -\! 1)N_{\rm BS}\! +\! 2,\! \cdots\! , kN_{\rm BS}\}$\;
\label{power1}
${\bm p}_k=[\|\widehat{\bm H}_{k,1} \|_F, \|\widehat{\bm H}_{k,2} \|_F,...,\|\widehat{\bm H}_{k,M} \|_F]^{\rm T}$~\%\ Calculate the channel matrix Frobenius norm of each subarray\;
\label{power2}
Acquire the set of subarrays $\Omega_k$ that have LoS links based on (\ref{Set_LoS})\;
\label{LoSet}
\If {$|\Omega_k|_c=1$}{
\label{BeginIf}
Set ${{\bm p}_k}_{[\Omega_k]}=0$ and $\Omega_k=\Omega_k \cup \{m~|~\text{max}_{m\in[M]} [{\bm p}_k]_m\}$~\%\ Avoid extreme conditions\; 
}
\label{EndIf}
\For{$m\in \Omega_k$}{
Perform EVD on ${\bm R}_{\widehat{\bm H}_{k,m}}$ and obtain the AoA estimation $\widehat{\theta}_{k,1,m}$ based on (\ref{OptTheta})\;
\label{EVD1}
Perform EVD on ${\bm R}_{\widehat{\bm H}^{\rm H}_{k,m}}$ and obtain the delay estimation $\widehat{\tau}_{k,1,m}$ based on (\ref{OptDelay})\;
\label{EVD2}
}
By collaboratively using parameters sensed by mutiple subarrays, we acquire $(\widehat{x}_k,\widehat{y}_k)$ based on (\ref{CoordEst})\;
\label{CoorEst}
}
{\bf Return}: $(\widehat{x}_k,\widehat{y}_k)$, $k\in{\cal K}$
\end{algorithm}

\subsubsection{Summarizing the Proposed MSCLoc Algorithm}
As shown in {\bf Algorithm \ref{ALG3}}, based on the estimated set of active UDs $\widehat{\cal S}$ and the estimated channel matrix ${\widehat{\bm H}}$, {\it{lines}}~$\it \ref{power1}$ and $\it \ref{power2}$ calculate the Frobenius norm (energy level) of the channel matrices corresponding to each subarray. Then, in {\it{line}}~$\it \ref{LoSet}$, the subarrays that have LoS links to the active UD are slected based on the energy levels. The threshold $\phi$ is set to 0.3. If one of LoS links has a very large energy level, {\it{line}}~$\it \ref{LoSet}$ might select only one subarray. To avoid this extreme condition, {\it{line}}~$\it \ref{BeginIf}\sim${\it{line}}~$\it \ref{EndIf}$ are performed. Furthermore, for each subarray with LoS link, {\it{line}}~$\it \ref{EVD1}$ and {\it{line}}~$\it \ref{EVD2}$ estimate the AoA $\widehat{\theta}_{k,1,m}$ and the path delay $\widehat{\tau}_{k,1,m}$ of the corresponding LoS link, respectively. Finally, in {\it{line}}~$\it \ref{CoorEst}$, by collaboratively using the AoAs and TDoAs estimated by multiple subarrays, we can acquire the estimated coordinates of each active UD, i.e., $(\widehat{x}_k,\widehat{y}_k)$, $k\in{\cal K}$.

\subsection{Computational Complexity Analysis}\label{S3.6}

{\color{black}
 The computational complexity of the proposed {\bf Algorithm \ref{ALG2}} mainly consists of the following three parts,
 where the notation $\mathcal{O}(N)$ stands for ``on order of $N$" with $N$ being the number of complex multiplications required at the $i$th iteration.
 \begin{itemize}
\item
 Calculate the correlations ({\it{line}}~$\it \ref{blockCorr}$): $\mathcal{O}(K N_{\rm BS} GM P)$;
\item
 Estimate the block channel vectors using LS ({\it{line}}~$\it \ref{LS}$): $\mathcal{O}( 2 i^2 N_{\rm s}^2 GMP\! +\! i^3 N_{\rm s}^3P\! +\! i N_{\rm s} GMP)$;
\item
 Update the residual vectors ({\it{line}}~$\it \ref{ResUpdate}$): $\mathcal{O}(GM i N_{\rm s} P)$.
\end{itemize}

Furthermore, the computational complexity of the proposed {\bf Algorithm \ref{ALG3}} mainly comes from the EVD processes in {\it{line}}~$\it \ref{EVD1}$ and {\it{line}}~$\it \ref{EVD2}$, which can be calculated as $\mathcal{O}(N_{\rm s}^3+P^3)$.
}

\section{Simulation Results}\label{S4}

In this section, an extensive simulation investigation is carried out to evaluate the performance of the proposed algorithms for AUD, CE, and localization in the XL-MIMO-based indoor scenario. The major simulation parameters are provided in Table~II. {\color{black}Note that we model the SNS property of XL-MIMO channel by randomly selecting $\widetilde{M}$ ($\widetilde{M}\leq M$) active subarrays for each path \cite{IimoriH_TWC22}, as indicated in (\ref{a_kpl}).} As shown in Fig.~\ref{FIG5}, we set $\overline{OA}\! =\! 1$\,m, $\overline{OB}\! =\! 3$\,m, $\overline{OD}\! =\! M\varDelta$, and $\overline{OC}\! =\! 3\overline{OD}/5$. The UDs and scatterers are randomly generated within the regions of $\{ (x,y)\vert 0\! \le\! x \! \le\! \overline{OD} \cup \overline{OB}\! \le\! y \! \le\! \overline{OC} \}$ and $\{ (x,y)\vert 0\! \le\! x \! \le\! \overline{OD} \cup \overline{OA}\! \le\! y \! \le\! \overline{OC} \}$, respectively. Without loss of generality, the LoS and NLoS links are generated randomly as well and there are at least two LoS links generated for each active UD. Then, the transmission distances and AoAs defined in Section~\ref{S2.2} are calculated based on the coordinates of the UDs and scatterers. Furthermore, the maximum transmission distance is considered as twice the length of $\overline{OD}$, hence we have the maximum delay spread $\tau_{\rm max}\! =\! 2\overline{OD}/v_{\rm c}$. Given $\tau_{\rm max}$ and the system bandwidth $f_{\rm s}$, the adjacent subcarrier interval and the coherence bandwidth can be calculated as $\varDelta f\! =\! f_{\rm s}/N_{\rm c}$ and $\varDelta B\! =\! 1/\tau_{\rm max}$, respectively \cite{MIMO-OFDM}. Hence, the number of intervals between adjacent pilot subcarriers is ${\bar P}\! =\! \left\lceil {\varDelta B}/{\varDelta f} \right\rceil$ and the number of pilot subcarriers is $P\! =\! \left\lceil N_{\rm c}/{\bar P} \right\rceil$ \cite{MIMO-OFDM}.

\begin{table*}[!t]
\vspace{-4mm}
\scriptsize
\centering
\captionsetup{font = {normalsize, color = {black}}, labelsep = period} 
\caption*{Table II: Simulation Parameter Settings \cite{mmW-GF-gao2,mmW-GF-shao1,IimoriH_TWC22,MYK,RicianChannel}}
\color{black}
\begin{threeparttable}
\begin{tabular}{|p{4.7cm}|p{1.1cm}||p{4.9cm}|p{2.2cm}|}
\hline
\hline
 {\textbf{Parameters}} & {\textbf{Values}}&{\textbf{Parameters}} & {\textbf{Values}} \\
\hline
 Carrier frequency $f_c$ & $28$\,GHz & Number of UDs & $K=60$ ($K_{\rm a}=6$)\\
\hline
System bandwidth\tnote{1}  $f_s$ & $200$\,MHz &  Number of active subarrays for each path & ${\cal U}[2,\lceil0.8M\rceil]$ \\
\hline
 Number of subarray (RFC) $M$& $10$ & Number of MPCs $L_k$ & ${\cal U}[1,5]$\\
\hline
 Number of antenna in each subarray $N_{\rm s}$ & $8$ & Rician factor $\gamma_k$ & $10$\\
\hline
 Adjacent subarray spacing $\varDelta$ & $5$\,m & Number of scatterers & ${\cal U}[5, 15]$\\
\hline
 Total number of subcarriers $N_{\rm c}$ & $2048$ & Background noise & $-174$\,dBm/Hz\\
\hline
\hline
\end{tabular}
\begin{tablenotes}
\scriptsize\item[1]{Due to the $200$\,MHz bandwidth, it is justifiable to neglect the frequency non-stationarity of mmWave channels \cite{FnS}.}
\end{tablenotes}
\end{threeparttable}
\end{table*}

To evaluate the AUD, CE, and localization performance of the proposed algorithms, we consider the following performance metrics, i.e., the AUD error probability $P_e$, the CE normalized mean square error (NMSE), and the root mean square error of coordinates estimation (RMSE$_{(x,y)}$),  which are respectively denoted by
\begin{align} 
{P_e} &= \big\| {\bm{\zeta}} - {\bm{\widehat \zeta}} \big\|_1 \big/ K , \label{Pe}\\ 
{\text{NMSE [dB]}} &= 10\log_{10}\left( \sum\limits_{p=1}^{P}\big\| {\bm h}_p - \widehat{{\bm h}_p} \big\|_2^2 \big/ \sum\limits_{p=1}^{P}\left\| {\bm h}_p \right\|_2^2 \right),\label{NMSE_dB}\\ 
\text{RMSE}_{(x,y)} &= \sqrt{\left(\left( \widehat{x}-x\right)^2+\left( \widehat{y}-y\right)^2\right)/ 2}. \label{RMSE}
\end{align}

{\color{black}
For comparison, we consider the following benchmarks. {\bf BSP}: BSP algorithm in \cite{DuY_WCL} with the number of active UDs $K_a$ known. {\bf BSAMP}: Block sparsity adaptive matching pursuit (BSAMP) algorithm in \cite{GaoZ_TSP15} with $K_a$ known. The adopted BSAMP algorithm stops its iterations if $K_a$ UDs are found out. {\bf BOMP}: BOMP algorithm in \cite{WangY_IoTJ22} with the noise variance $\epsilon_{\rm th}$ known, where  the entire XL-MIMO antenna array is considered as a block. {\bf BOMP-SA}: BOMP-subarray (BOMP-SA) considers each subarray as a block as in the proposed StrBOMP algorithm. Specifically, BOMP-SA performs {\it{lines}}~$\it \ref{A1_Init}-\ref{ActSet}$ and {\it{lines}}~$\it \ref{ActIndic}-\ref{A1_end}$ of the proposed Algorithm 1. {\bf Oracle LS}: LS detector with perfectly known active UDs as in \cite{TWCqiao}. {\bf Oracle LS-SA}: Oracle LS-subarray (Oracle-LS-SA) uses LS detector with perfectly known active subarrays related to the active UDs. In the simulations, we will first verify the performance of the proposed StrBOMP algorithm and the benchmarks with the same MMV sensing matrices. Then, we will show that the proposed StrBOMP algorithm with GMMV sensing matrices (denoted as ``Proposed (GMMV)") performs much better than that with MMV sensing matrix (denoted as ``Proposed (MMV)"). The computational complexity of the proposed StrBOMP algorithm and the benchmarks are shown in Table III. It can be seen that the proposed StrOMP algorithm has the same order of computational complexity as the BOMP-SA and BOMP algorithms. While, the BSP and BSAMP algorithms have much larger computational complexity.

\begin{table*}[!t]
\vspace{-2mm}
\scriptsize
\centering
\captionsetup{font = {normalsize, color = {black}}, labelsep = period} 
\caption*{Table III: Computational complexity comparison of different algorithms for joint AUD and CE}
\color{black}
\begin{threeparttable}
\begin{tabular}{|p{2.3cm}|p{9.3cm}|p{2cm}|}
\Xhline{1.2pt}
\makecell[c]{\bf Algorithms} & \makecell[c]{\bf Computational complexity}&\makecell[c]{\bf Values ($10^{10}$)}\\
\Xhline{1.2pt}
\makecell[c]{BSAMP} & \makecell[c]{$\mathcal{O}\{2K_{\rm a}KN_{\rm BS}GMP+\sum\nolimits_{i=1}^{K_{\rm a}}[11 Pi^3N_{\rm BS}^3+GMP(14i^2N_{\rm BS}^2+10iN_{\rm BS})]\}$}& \makecell[c]{44.2}\\
\hline
\makecell[c]{BSP} &\makecell[c]{$\mathcal{O}\{2KN_{\rm BS}GMP+\sum\nolimits_{i=1}^{2}[Pi^3K_{\rm a}^3N_{\rm BS}^3+2GMP(i^2K_{\rm a}^2N_{\rm BS}^2+iK_{\rm a}N_{\rm BS})]\}$}& \makecell[c]{14.4}\\
\hline
\makecell[c]{BOMP}&\makecell[c]{$\mathcal{O}\{K_{\rm a}KN_{\rm BS}GMP+\sum\nolimits_{i=1}^{K_{\rm a}}[Pi^3N_{\rm BS}^3+2GMP(i^2N_{\rm BS}^2+iN_{\rm BS})]\}$}& \makecell[c]{5.5}\\
\hline
\makecell[c]{BOMP-SA\tnote{1}} &\makecell[c]{$\mathcal{O}\{IKN_{\rm BS}GMP+\sum\nolimits_{i=1}^{I}[Pi^3N_{\rm s}^3+2GMP(i^2N_{\rm s}^2+iN_{\rm s})]\}$ }& \makecell[c]{5.3}\\
\hline
\makecell[c]{Proposed StrBOMP\\(MMV or GMMV)} &\makecell[c]{$\mathcal{O}\{IKN_{\rm BS}GMP+\sum\nolimits_{i=1}^{I}[Pi^3N_{\rm s}^3+2GMP(i^2N_{\rm s}^2+iN_{\rm s})]\}$ }& \makecell[c]{5.3}\\
\Xhline{1.2pt}
\end{tabular}
\begin{tablenotes}
\scriptsize\item[1]Due to the near-field SNS property, $I\leq K_{\rm a}M$. For the values in the third column, we use parameters from Table II and set $I=\left\lceil {K_{\rm a}M}/{2} \right\rceil$.
\end{tablenotes}
\end{threeparttable}
\vspace{-6mm}
\end{table*}

\begin{figure*}[t]
\centering
\subfigure[]{
    \begin{minipage}[t]{0.5\linewidth}
        \centering
        \includegraphics[width=0.8\columnwidth]{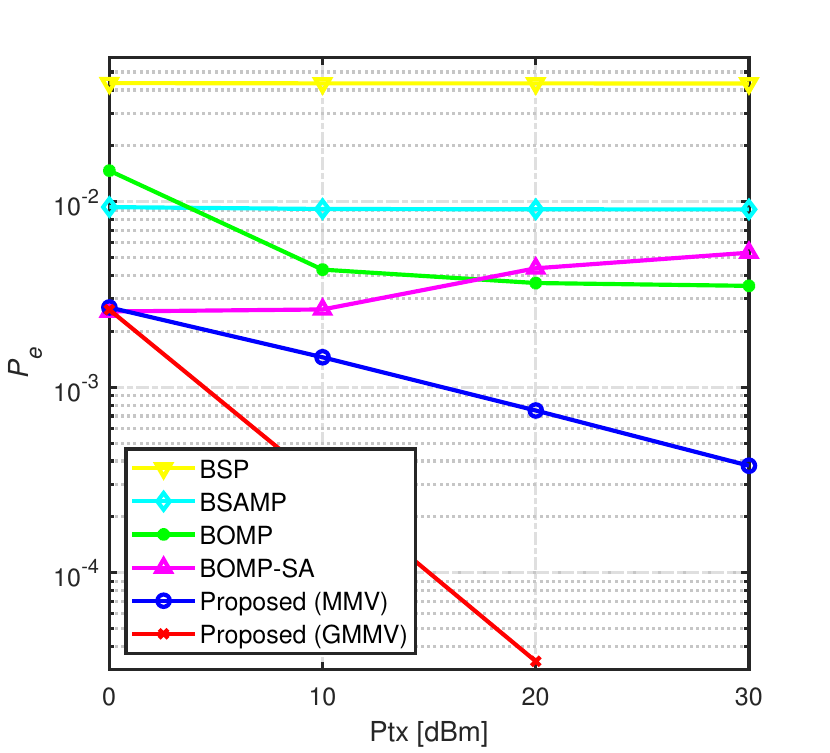}\\
\label{FIG6(a)}
    \end{minipage}%
}%
\subfigure[]{
    \begin{minipage}[t]{0.5\linewidth}
        \centering
        \includegraphics[width=0.8\columnwidth]{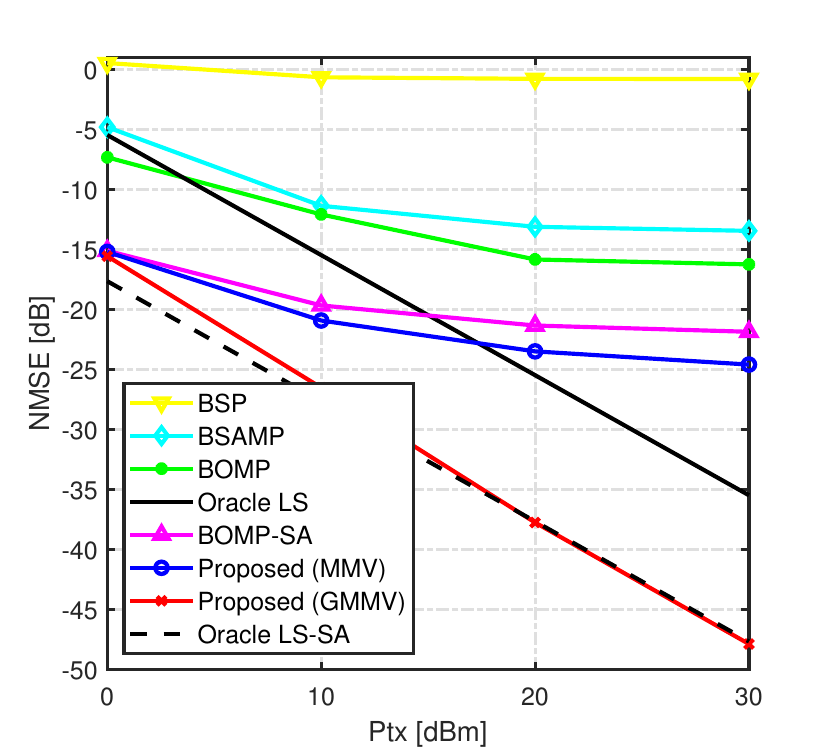}\\
\label{FIG6(b)}
    \end{minipage}%
}%
\centering
\setlength{\abovecaptionskip}{-1mm}
\captionsetup{font={footnotesize, color = {black}}, singlelinecheck = off, justification = raggedright,name={Fig.},labelsep=period}
\caption{Performance comparison of different AUD and CE schemes against the transmit power $P_{\rm tx}$ with time overhead $G\! =\! 50$: (a) $P_e$ performance of AUD; (b) NMSE performance of CE.}
\label{FIG6}
\end{figure*}

Fig.~\ref{FIG6} compares the AUD and CE performance of different algorithms against the transmit power $P_{\rm tx}$ of each UD with the time overhead $G\! =\! 50$. {\color{black} 
Owing to the exploitation of the near-field SNS property, it can be seen that the proposed StrBOMP algorithm outperforms the BSP, BSAMP, BOMP, and BOMP-SA algorithms in terms of both AUD and CE performance. Furthermore, due to the diversity gain achieved by diversifying sensing matrices, the ``Proposed (GMMV)" algorithm achieves much better AUD and CE performance than the ``Proposed (MMV)" algorithm. Note that the BOMP-SA algorithm also exploits the SNS property, i.e., considering each subarray as a block. However, the ignorance of XL-MIMO structure results in its poor AUD performance. In addition, because of the SNS property, i.e., only part of the subarrays are active for each active UD, Oracle LS-SA algorithm is better than Oracle LS algorithm. Also, it is shown from Fig.~\ref{FIG6(b)} that the ``Proposed (GMMV)" algorithm can approach the CE performance of the Oracle LS-SA algorithm.
}

Fig.~\ref{FIG7} compares the AUD and CE performance of different schemes against time overhead $G$ at transmit power $P_{\rm tx}\! =\! 20$\,dBm. 
{\color{black}From Fig.~\ref{FIG7(b)}, as the time overhead increases, the CE performance of the proposed StrOMP and BOMP-SA algorithms can approach the Oracle LS-SA lower bound. While, the BOMP and BSAMP algorithms approach the Oracle LS lower bound. Again, it is shown that the proposed StrBOMP algorithm outperforms all the benchmarks, indicating that the proposed StrBOMP algorithm can significantly reduce the time overhead/latency. Furthermore, it can be seen that the BOMP algorithm performs better than the BSP and BSAMP algorithms. The implies that the block structure of the entire XL-MIMO array helps BOMP algorithm to find the correct active UD in every iteration. However, with limited number of observations, the support pruning steps in both BSP and BSAMP algorithms may fail to prune the correct UDs, resulting in the poor performance. Note that if $G\leq 40$, the number of observations is less than the number of unknown parameters, i.e., it is an underdetermined problem. Hence, the NMSE performance of the algorithms under $G\leq 40$ seems irregular.
}

\begin{figure*}[t]
\centering
\subfigure[]{
    \begin{minipage}[t]{0.5\linewidth}
        \centering
        \includegraphics[width=0.8\columnwidth]{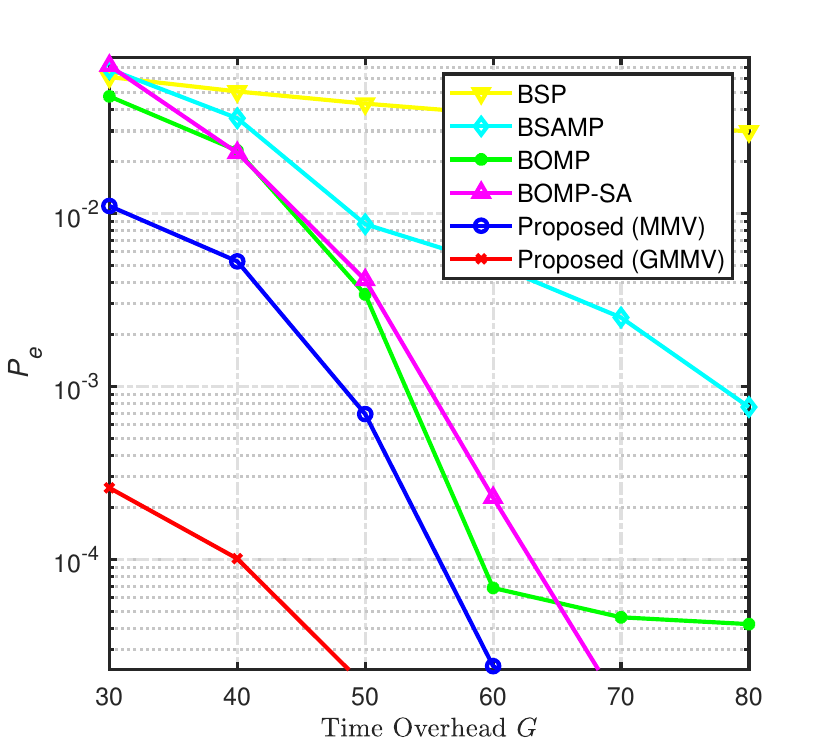}\\
\label{FIG7(a)}
    \end{minipage}%
}%
\subfigure[]{
    \begin{minipage}[t]{0.5\linewidth}
        \centering
        \includegraphics[width=0.8\columnwidth]{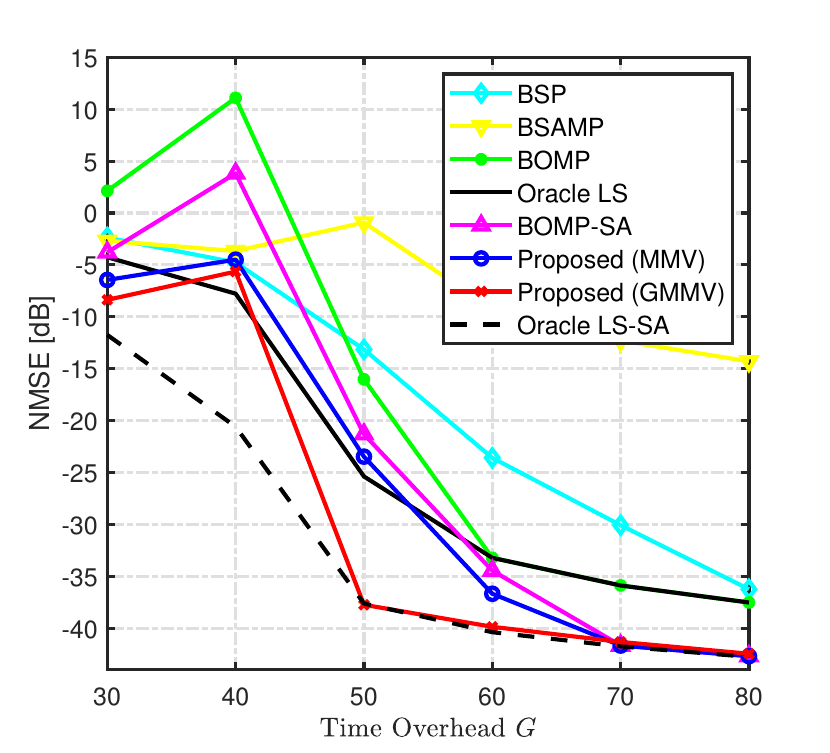}\\
\label{FIG7(b)}
    \end{minipage}%
}%
\centering
\setlength{\abovecaptionskip}{-1mm}
\captionsetup{font={footnotesize, color = {black}}, singlelinecheck = off, justification = raggedright,name={Fig.},labelsep=period}
\caption{Performance comparison of different AUD and CE schemes against time overhead $G$ at transmit power $P_{\rm tx}\! =\! 20$\,dBm: (a) $P_e$ performance of AUD; (b) NMSE performance of CE.}
\label{FIG7}
\end{figure*}

 \begin{figure*}[t]
\centering
\subfigure[]{
    \begin{minipage}[t]{0.5\linewidth}
        \centering
        \includegraphics[width=0.8\columnwidth]{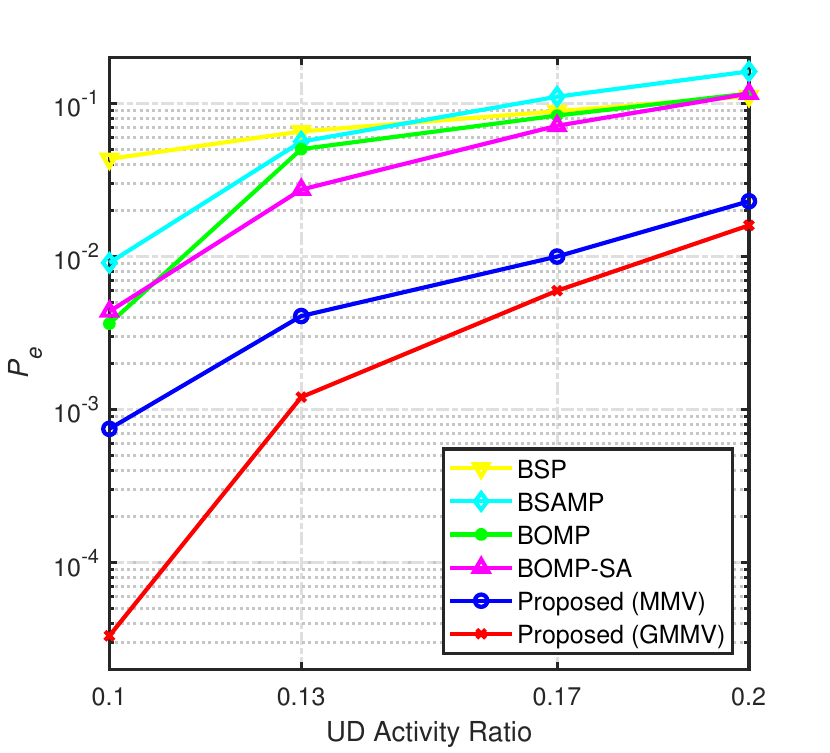}\\
\label{FIG8(a)}
    \end{minipage}%
}%
\subfigure[]{
    \begin{minipage}[t]{0.5\linewidth}
        \centering
        \includegraphics[width=0.8\columnwidth]{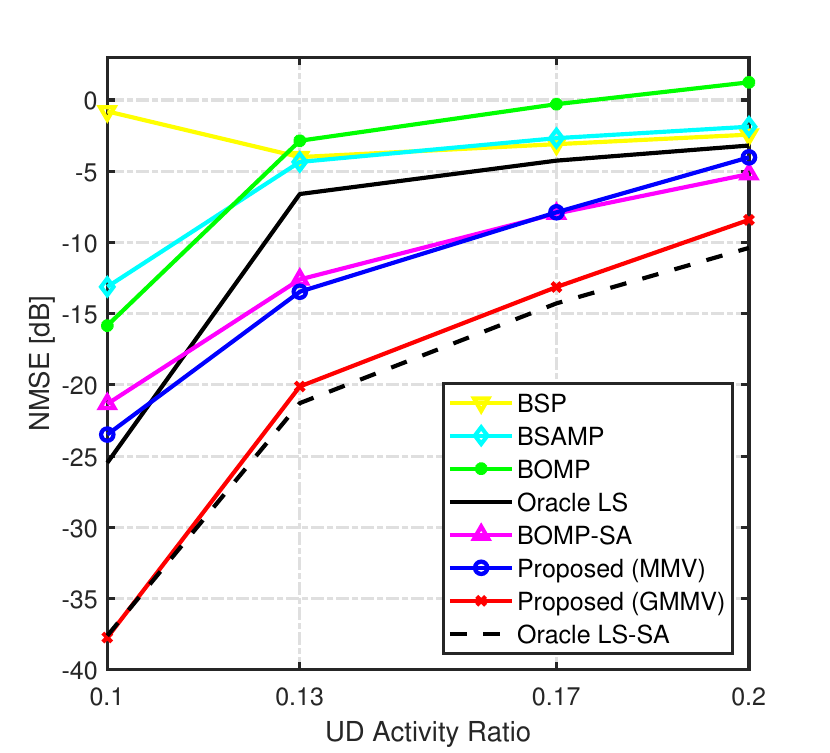}\\
\label{FIG8(b)}
    \end{minipage}%
}%
\centering
\setlength{\abovecaptionskip}{-1mm}
\captionsetup{font={footnotesize, color = {black}}, singlelinecheck = off, justification = raggedright,name={Fig.},labelsep=period}
\caption{Performance comparison of different AUD and CE schemes against UD activity ratio ($K_{\rm a}/K$) with transmit power $P_{\rm tx}\! =\! 20$\,dBm, time overhead $G=50$, and fixed $K=60$: (a) $P_e$ performance of AUD; (b) NMSE performance of CE. }
\label{FIG8}
\end{figure*}

 To examine the effectiveness of the proposed algorithm under different sparsity levels, Fig.~\ref{FIG8} compares the $P_e$ and NMSE performance of different algorithms against UD activity ratio, i.e., $K_{\rm a}/K$, with transmit power $P_{\rm tx}\! =\! 20$\,dBm, time overhead $G=50$, and fixed $K=60$. {\color{black}It can be seen from Fig.~\ref{FIG8} that although the $P_e$ performance of all the algorithms degrades with the increase of UD activity ratio, the ``Proposed (GMMV)" algorithm is still significantly superior to all the benchmarks. In addition, the ``Proposed (GMMV)" algorithm is close to the Oracle LS-SA lower bound under different UD activity ratios.}

\begin{figure*}[t]
\centering
\subfigure[]{
    \begin{minipage}[t]{0.5\linewidth}
        \centering
        \includegraphics[width=0.8\columnwidth]{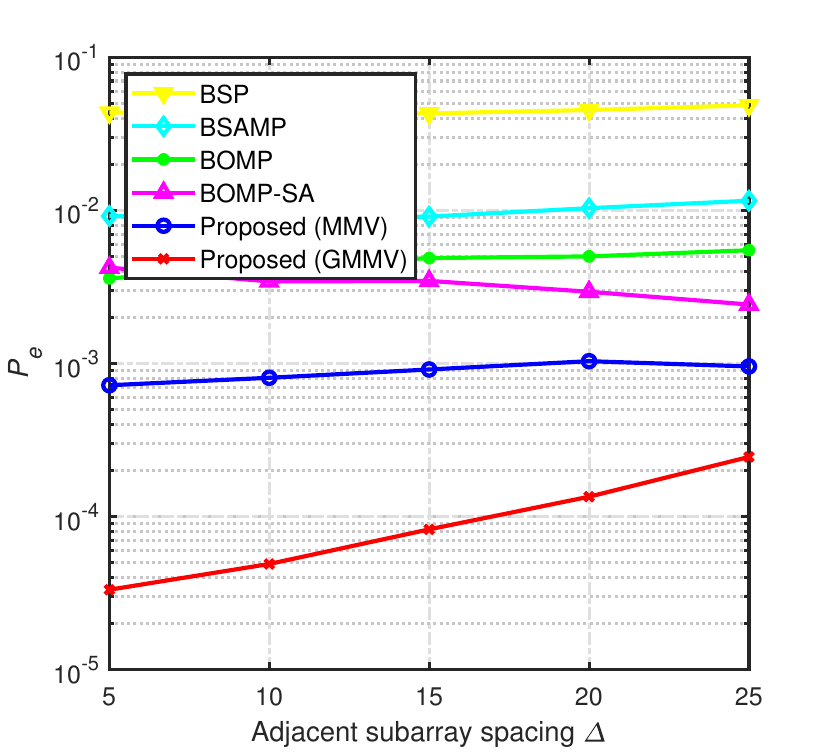}\\
\label{FIG9(a)}
    \end{minipage}%
}%
\subfigure[]{
    \begin{minipage}[t]{0.5\linewidth}
        \centering
        \includegraphics[width=0.8\columnwidth]{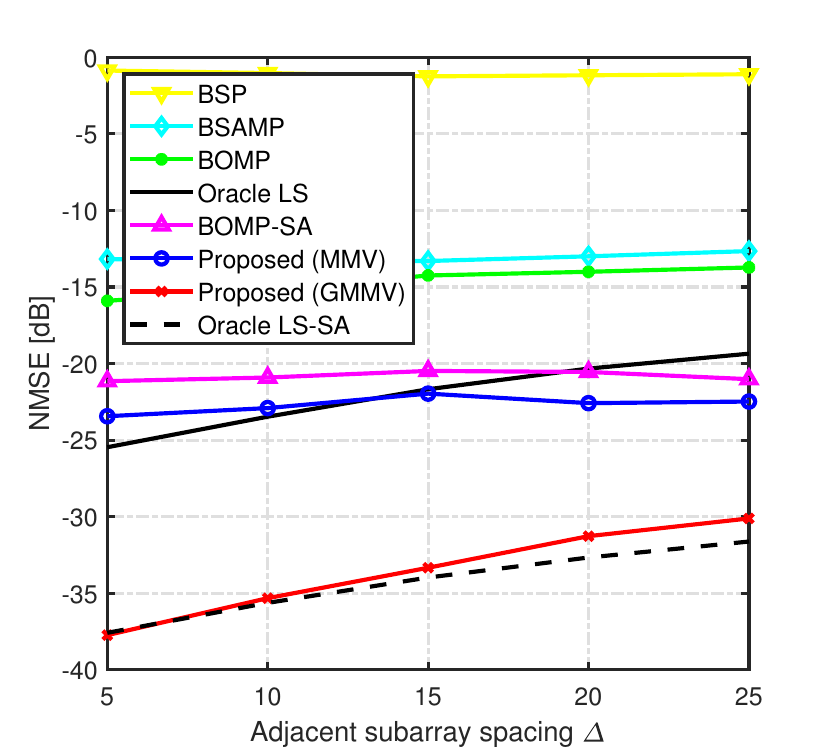}\\
\label{FIG9(b)}
    \end{minipage}%
}%
\centering
\setlength{\abovecaptionskip}{-1mm}
\captionsetup{font={footnotesize, color = {black}}, singlelinecheck = off, justification = raggedright,name={Fig.},labelsep=period}
\caption{Performance comparison of different AUD and CE schemes against adjacent subarray spacing $\varDelta$ with the transmit power $P_{\rm tx}\! =\! 20$\,dBm and the time overhead $G\! =\! 50$: (a) $P_e$ performance of AUD; (b) NMSE performance of CE. }
\label{FIG9}
\end{figure*}

 To illustrate the impact of the channel conditions in the indoor environment (shown in Fig. \ref{FIG5}), Fig.~\ref{FIG9} compares the AUD and NMSE performance of different algorithms against adjacent subarray spacing $\varDelta$ with the transmit power $P_{\rm tx}\! =\! 20$\,dBm and the time overhead $G\! =\! 50$. To alter the near-field environment, we only change the adjacent subarray spacing $\varDelta$. 
{\color{black}It can be seen that our proposed StrBOMP algorithm always considerably outperform the other benchmark algorithms with the increase of the indoor space. Also, due to the severer large scale fading as $\varDelta$ increases, the proposed StrBOMP algorithm has a slight $P_e$ and NMSE performance degradation. To compensate for this performance loss, we can increase the transmit power of UDs for better performance if the service area is very large.
}
 
 \begin{figure*}[t]
\centering
\subfigure[]{
    \begin{minipage}[t]{0.5\linewidth}
        \centering
        \includegraphics[width=0.8\columnwidth]{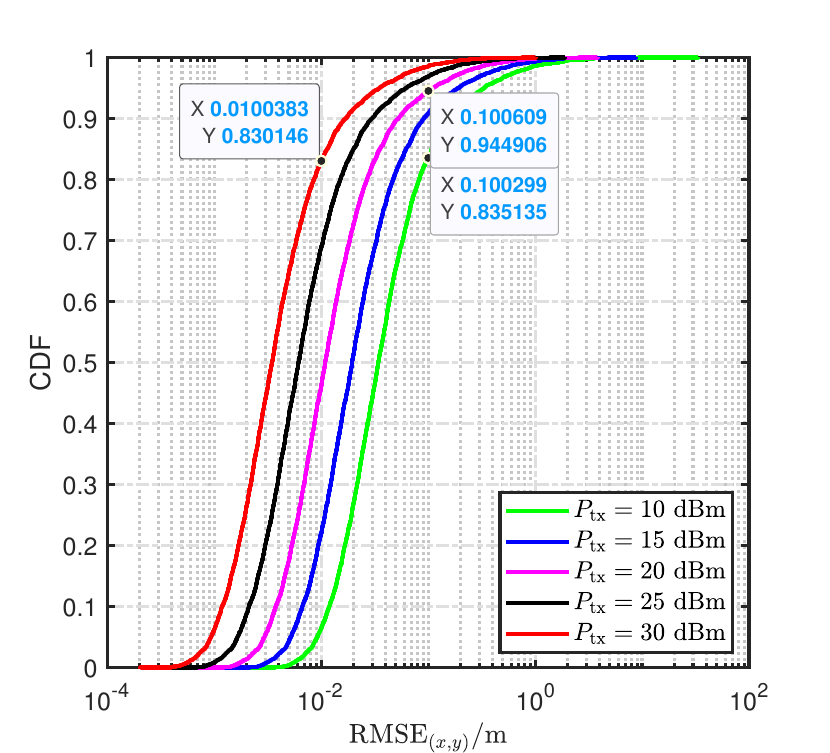}\\
\label{FIG10(a)}
    \end{minipage}%
}%
\subfigure[]{
    \begin{minipage}[t]{0.5\linewidth}
        \centering
        \includegraphics[width=0.8\columnwidth]{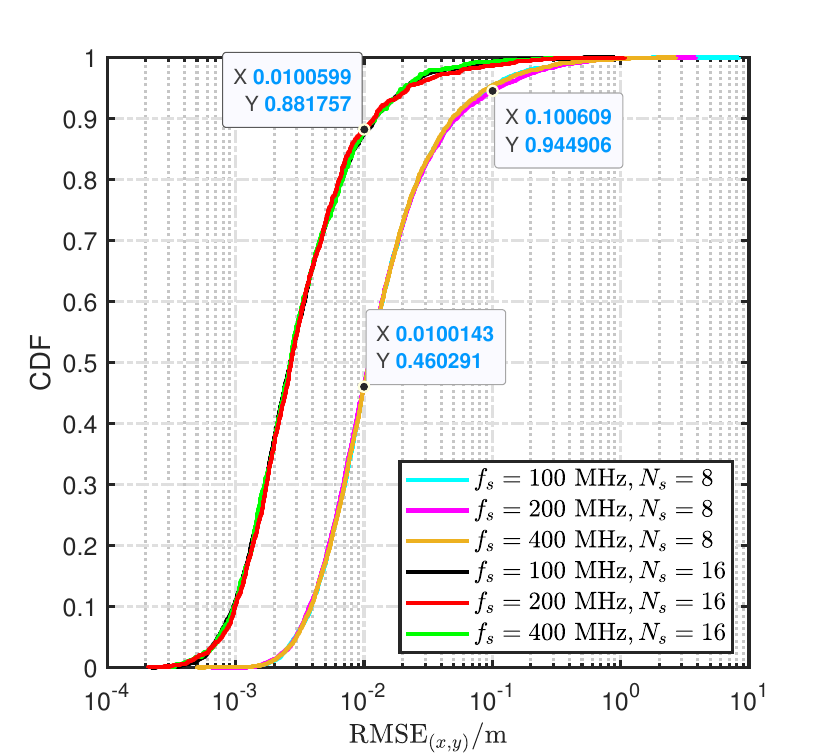}\\
\label{FIG10(b)}
    \end{minipage}%
}%
\centering
\setlength{\abovecaptionskip}{-1mm}
\captionsetup{font={footnotesize, color = {black}}, singlelinecheck = off, justification = raggedright,name={Fig.},labelsep=period}
\caption{Relation between CDF and RMSE$_{(x,y)}$, where the time overhead $G\! =\! 50$ for $N_{\rm s}=8$ and $G\! =\! 100$ for $N_{\rm s}=16$: (a) CDF of RMSE$_{(x,y)}$ under various transmit powers $P_{\rm tx}$, where $f_s=200$\,MHz and $N_{\rm s}=8$; (b) CDF of RMSE$_{(x,y)}$ under various bandwidth $f_s$ and number of antennas of each subarray $N_{\rm s}$, where $P_{\rm tx}=20$\,dBm.}
\label{FIG10}
\end{figure*}

To investigate the localization accuracy of the proposed MSCLoc algorithm, the cumulative distribution function (CDF) of RMSE$_{(x,y)}$ is shown in Fig. \ref{FIG10}. The proposed MSCLoc algorithm is sequentially executed after the proposed StrBOMP (GMMV sensing matrices) algorithm,with the time overhead $G\! =\! 50$ for $N_{\rm s}=8$ and $G\! =\! 100$ for $N_{\rm s}=16$. The CDFs of RMSE$_{(x,y)}$ under different transmit power $P_{\rm tx}$ are plotted in Fig. \ref{FIG10(a)}. It can be observed that the RMSE$_{(x,y)}$ of the cascade algorithm (i.e., proposed StrBOMP and MSCLoc algorithm) can be lower than centimeter (i.e., $10^{-2}$\,m) with probability 83\% at transmit power 30\,dBm. As the transmit power decreases, the RMSE$_{(x,y)}$ decreases as well. However, the localization accuracy at $P_{\rm tx}=10$\,dBm still has decimeter-level accuracy with probability 83.5\%. Fig. \ref{FIG10(b)} compares RMSE$_{(x,y)}$ under different bandwidth $f_s$ and the number of antennas of each subarray $N_{\rm s}$, where $P_{\rm tx}$ is fixed to 20\,dBm. Due to the enhanced accuracy of AoAs estimation, it can be observed that by increasing $N_{\rm s}$ from $8$ to $16$, the probability of RMSE$_{(x,y)}\le10^{-2}$ increases from 46\% to 88\%. {\color{black}While, if we fix $N_{\rm s}$, the curves of $f_s=100$\,MHz, $f_s=200$\,MHz and $f_s=400$\,MHz are nearly coincided, which indicates that the TDoAs are well estimated for $f_s\geq100$. Thanks to the mmWave XL-MIMO, by using the proposed algorithms at the BS, joint AUD and CE as well as centimeter-level UD localization can be realized efficiently.}

\vspace{-3mm}
\section{Conclusions}\label{S5}
To support indoor massive IoT access with low latency, high data rate and high localization accuracy, this paper proposed a mmWave XL-MIMO-based grant-free massive access scheme. The mmWave XL-MIMO array adopted widely located multiple subarrays to guarantee LoS links. We modeled the XL-MIMO-based massive access problem considering the near-field SNS property of the channel. {\color{black}By exploiting this SNS property and the block sparsity of subarrays, we proposed a StrBOMP algorithm for efficient joint AUD and CE. In addition, the structure of XL-MIMO is used to refine the AUD and CE results for better performance.} After estimating the XL-MIMO channel, an MSCLoc algorithm was proposed for localization, where the AoAs and TDoAs between active UDs and related subarrays were extracted based on the MUSIC algorithm. Then, these parameters were jointly utilized to acquire the coordinates of active UDs. Finally, simulation results verified that the proposed algorithms outperform the cutting-edge schemes in terms of AUD and CE performance, and can achieve centimeter-level localization accuracy. {\color{black}Furthermore, there are numerous unexplored avenues for future research. These include emerging challenges such as the development of XL-MIMO-based near-field joint sensing and beamforming design, the utilization of IRS/THz for near-field multiple access and wireless sensing, and the design of data/model-driven deep learning algorithms as novel approaches.}

\ifCLASSOPTIONcaptionsoff
  \newpage
\fi

\end{document}